%% file: main.tex
\documentclass[12pt, a4paper]{article}
\usepackage{fullpage,times}

\AtBeginDocument{%
  \providecommand\BibTeX{{%
    \normalfont B\kern-0.5em{\scshape i\kern-0.25em b}\kern-0.8em\TeX}}}
 
\usepackage{textcomp}
\usepackage{comment}
\usepackage[numbers]{natbib}
\usepackage{multicol} 
\usepackage{authblk}
\usepackage{mathtools}
\usepackage[export]{adjustbox}
\usepackage{verbatim}
\usepackage{tcolorbox}
\usepackage{url}
\usepackage{mdframed}
\usepackage{amssymb}
\usepackage{hyperref}
\usepackage{multicol}
\usepackage{caption}
\usepackage{subcaption}
\usepackage{multirow}
\usepackage[linesnumbered,ruled,vlined,resetcount,noend]{algorithm2e}
\usepackage{fancyhdr}
\usepackage{amsmath}
\usepackage{amssymb}
\usepackage{float}
\usepackage{graphicx}
\usepackage{tabularx}
\usepackage{enumitem}
\usepackage{makecell}

\usepackage{lipsum}
\usepackage{subcaption}
\usepackage{pifont}
\usepackage{enumitem}
\renewcommand*{\listalgorithmcfname}{List of Protocols}
\renewcommand*{\algorithmcfname}{Protocol}
\renewcommand*{\algorithmautorefname}{protocol}

\newcounter{protocol}

\newcommand{\VKalice}{\mathsf{VK_{Alice}}}

\newcommand{\rh}{\mathrm{RH}}
\newcommand{\cursucc}{\mathsf{curr_{succ}}}
\newcommand{\amount}{\mathsf{amt}}
\newcommand{\bcwrite}{\mathsf{BC.Write}}
\newcommand{\SK}{\mathsf{SK}}
\newcommand{\VK}{\mathsf{VK}}
\newcommand{\alice}{\mathsf{Alice}}
\newcommand{\bob}{\mathsf{Bob}}
\newcommand{\denise}{\mathsf{Denise}}

\newcommand{\sk}{\mathsf{sk}}
\newcommand{\vk}{\mathsf{vk}}
\newcommand{\pid}{\mathsf{pathid}}
\newcommand{\send}{\mathsf{sender}}
\newcommand{\rec}{\mathsf{receiver}}
\newcommand{\curi}{\mathsf{i}}
\newcommand{\curp}{\mathsf{p}}

\newcommand{\erh}{\mathit{endRH}}
\newcommand{\nearrh}{\mathit{nearRH}} 
\newcommand{\key}{\mathsf{key}}
\newcommand{\hc}{\mathsf{hc}}
\newcommand{\ripd}{\mathit{RACED}}
\newcommand{\txid}{\mathit{txid}}
\newcommand{\curl}{\mathsf{L}}
\newcommand{\curj}{\mathsf{J}}
\newcommand{\curs}{\mathsf{S}}
\newcommand{\curtime}{\mathsf{curr_{time}}}
\newcommand{\maxamount}{\mathsf{max}}
\newcommand{\curprev}{\mathsf{{previous}}}
\newcommand{\curnext}{\mathsf{{next}}}
\newcommand{\verify}{\mathsf{Verify}}
\newcommand{\curk}{\mathsf{{k}}}

\newcommand{\txtable}{\mathsf{{Txtable}}}
\newcommand{\hset}{\mathbb{H}}
\newcommand{\dset}{\mathbb{D}}
\newcommand{\iset}{\mathbb{I}}

 \newcommand{\pset}{\mathbb{P}}
\newcommand{\lset}{\mathcal{L}}
\newcommand{\tset}{\mathbb{T}}
\newcommand{\wset}{\mathbb{W}}
\newcommand{\bset}{\mathcal{B}}
\newcommand{\jset}{\mathcal{J}}
\newcommand{\mset}{\mathcal{M}}

\newcommand{\add}{\mathsf{Add}}

\newcommand{\delete}{\mathsf{Delete}}
\newcommand{\open}{\mathsf{PC.Open}}
\newcommand{\close}{\mathsf{PC.Close}}

\newcommand{\lookup}{\mathsf{FT.Lookup}}
\newcommand{\succlookup}{\mathsf{Succ.Lookup}}
\newcommand{\sign}{\mathsf{Sign}}
\newcommand{\search}{\mathsf{FT.Search}}
\newcommand{\htlcpay}{\mathsf{HTLC.Pay}}

\newcommand{\retrieve}{\mathsf{FT.Retrieve}}
\newcommand{\retrievenext}{\mathsf{RetrieveNext}}
\newcommand{\pop}{\mathsf{Pop}}
\newcommand{\removeduplicate}{\mathsf{RemoveDuplicates}}
\newcommand{\ftcompute}{\mathsf{FT.Compute}}
\newcommand{\retrieveneighbors}{\mathsf{RetrieveNeighbors}}

\newcommand{\push}{\mathsf{Push}}

\newcommand{\simulator}{\mathcal{S}}
\newcommand{\adversary}{\mathcal{A}}

\newcommand{\ffindpath}{\mathcal{F}_{Findpath}}
\newcommand{\fsigverify}{\mathcal{F}_{sig}}

\newcommand{\keygenLT}{\mathsf{KeyGen}}

\newcommand{\longverificationkey}{\mathsf{Long \ Term \ Verification \ Key}}
\newcommand{\tempverificationkey}{\mathsf{Temporary \ Verification \ Key}}
\newcommand{\signature}{\mathsf{Signature}}
\newcommand{\faux}{\mathcal{F}_{aux}}
\newcommand{\fhtlc}{\mathcal{F}_{htlc}}
\newcommand{\fdht}{\mathcal{F}_{DHT}}

\newcommand{\fingertable}{\mathsf{FingerTable}}
\newcommand{\idsender}{sid}

\newcommand{\finit}{\mathcal{F}_{init}}
\newcommand{\neighbors}{\mathsf{Immediate~Neighbors}}

\newcommand{\verification}{\mathsf{Verify}}
\newcommand{\utable}{\mathit{utable}}

\newcommand{\rtable}{\mathit{rtable}}
\newcommand{\fprime}{f ^ \prime}

\newcommand{\fpayment}{\mathcal{F}_{Payment}}
\newcommand{\honest}{\mathsf{Honest}}
\newcommand{\dishonest}{\mathsf{Dishonest}}
\newcommand{\fraced}{\mathsf{F_{RACED}}}
\newcommand{\fpcn}{\mathcal{F_{PCN}}}
\usepackage{amsthm}
\newtheorem{theorem}{Theorem}[section]

\newtheorem{definition}{Definition}[section]

\makeatletter
\newcommand\tabcaption{\def\@captype{table}\caption}
\newcommand\figcaption{\def\@captype{figure}\caption}
\makeatother

\newcommand{\xmark}{\text{\ding{55}}}
\usepackage [
n ,
advantage ,
operators,
sets,
adversary ,
landau ,
probability ,
notions, 
logic,
ff,
mm,
primitives,
events,
complexity,
asymptotics ,
keys
]{cryptocode}
 \newlength\myindent
\makeatletter
\def\@copyrightspace{\relax}
\makeatother
\begin{document}

 \title{RACED: \underline{R}outing in P\underline{A}yment \underline{C}hannel N\underline{E}tworks Using \underline{D}istributed Hash Tables\footnote{A short version of this work has been accepted to the $19^{th}$ ACM ASIA Conference on Computer and Communications Security (ACM ASIACCS 2024). This material is based upon work supported by the National Science Foundation under Award No. 2148358, 1914635, and the Department of Energy. Any opinions, findings and conclusions or recommendations expressed in this material are those of the authors and do not necessarily reflect the views of the National Science Foundation and the Department of Energy.}}


 \author{Kartick Kolachala, Mohammed Ababneh, Roopa Vishwanathan}
 \affil{New Mexico State University, Las Cruces, NM, USA}
 \affil{\href{mailto:kart1712@nmsu.edu}{kart1712@nmsu.edu}, \href{mailto:mababneh@nmsu.edu@nmsu.edu}{mababneh@nmsu.edu}, \href{mailto:roopav@nmsu.edu}{roopav@nmsu.edu}}





  \maketitle
 \begin{abstract}
The Bitcoin scalability problem has led to the development of off-chain financial mechanisms such as payment channel networks (PCNs) which help users process transactions of varying amounts, including micro-payment transactions, without writing each transaction to the blockchain. Since PCNs only allow path-based transactions, effective, secure routing protocols that find a path between a sender and receiver are fundamental to PCN operations. In this paper, we propose $\ripd$, a routing protocol that leverages the idea of  Distributed Hash Tables (DHTs) to route transactions in PCNs in a fast and secure way. 
Our experiments on real-world transaction datasets show that $\ripd$ gives an average transaction success ratio of 98.74\%, an average pathfinding time of 31.242 seconds, which is $1.65 \times 10^3$, $1.8 \times 10^3$, and $4 \times 10^2$ times 
faster than three other recent routing protocols that offer comparable security/privacy properties. We rigorously analyze and prove the security of $\ripd$ in the Universal Composability framework.

\end{abstract}

 \pagestyle{plain}


\input{introduction_full}
\input{related_full}

\input{sysmod_full}

\input{advmod_full}

\input{construction_full}
\input{algo_full}
\input{impl_full}

\input{secanalysis_full}

\input{conc_full}


\bibliographystyle{plain}
\bibliography{references_full}
   \input{appendix_full}

\end{document}

%% file: introduction_full.tex
\section{Introduction}
\label{sec:intro}
The development of cryptocurrencies, which began with the Bitcoin white paper~\cite{nakamoto2008bitcoin} in 2009, has disrupted banking and financial processes across the globe. 
As of February 2023, Bitcoin's market capitalization stands at 453 Billion USD ~\cite{Cryptostats}. However, the throughput of transactions involving cryptocurrencies is extremely low due to the high latency of transaction confirmation on the blockchain. For instance, the transaction processing speed of Bitcoin is 5-7 transactions per second and that of Ethereum is 15-30 transactions per second~\cite{btctx,bitkan,blockchair,binance}.
This is in sharp contrast with traditional fiat currency's throughput, e.g., Visa processes over 65,000 transactions per second~\cite{visa}. 
One of the most promising solutions to this problem is off-chain payment channels. Two parties create a payment channel on a blockchain with some initial balance, following which they can send an unlimited number of payments to each other using that channel without writing anything to the blockchain. 
Access to the blockchain is only needed either if there is a dispute or the two parties involved decide to close the channel. 
 
\par
This idea can be extended to enable transactions between two parties that may not have a
payment channel currently open between them. Decentralized payment channel networks (PCNs) that enable
transitive payments have been proposed such as \cite{lightning,Malavolta2016SilentWhispersES,roos2017settling,sprites,flare},  where two unconnected users can send/receive
payments if there exists a path comprising of several users with payment channels between them.
The first such network was the Lightning Network, which operates on top of the Bitcoin blockchain \cite{lightning}. 
Lately, Lightning Network has become one of the fastest-growing PCNs. 
Between January 2021 and December 2021, there were a total of 28 Million unique channels opened in the Lightning Network, with an average of 73,733 new channels created every day.  The number of unique nodes (unique public key pairs) involved in channel opening during this period was 6.5 Million \cite{btcpay}. The market capitalization of Lightning Network is USD 1 Million as of 2023.
Several other payment channel networks and credit networks have been developed, which have later evolved into blockchain-based decentralized financial ecosystems,  such as Ripple \cite{ripple}, which has a current market value of 20 Billion USD \cite{ripplemarket}, (increased from 9.97 Billion USD in 2017 and peaked at 64 Billion USD in April 2021) and Stellar \cite{stellar}.
Between January 2021 to December 2021, there were a total of 15 Million transactions recorded on the Ripple ledger, with an average of 1 Million transactions recorded every month~\cite{rippleapi}. These numbers indicate the size and growth of PCNs. 

A major advantage of PCNs is that they facilitate micro-payments between users that can be as small as $10^{-7}$ BTC \cite{minlightning}. Apart from this, the fees charged by PCNs to route payments are a fraction of the on-chain transaction fees charged by the underlying blockchain. The problem of finding an efficient route between a sender and receiver in a PCN is challenging and has attracted considerable attention from the research community ~\cite{blanc,Malavolta2016SilentWhispersES,lightpir,flash,vein}.
While there have been many elegant routing protocols developed recently for PCNs, each one comes with its own set of limitations. Some routing protocols do not provide security of transactions nor privacy of the users~\cite{roos2017settling,flash,coinexpress}, while others do not support concurrent transactions \cite{Malavolta2016SilentWhispersES,sivaraman2020high,flash,coinexpress,FSTR}. Some routing protocols need trusted entities to route payments~\cite{Malavolta2016SilentWhispersES}, while others implement source routing, in which the network topology needs to be known to all nodes~\cite{flash}. 
In this paper, we present a novel routing mechanism called $\ripd$ that uses distributed hash tables (DHT) to route the payment from the sender to the receiver in PCNs.

Current routing protocols for PCNs traverse the entire network in the worst case to route a payment from a sender to a receiver, with a maximum path length of $n-1$ hops, in a network of $n$ nodes. Using DHTs will help us in reducing this path length since the complexity of locating any node in a DHT is logarithmic in the number of nodes in the DHT. This consequently reduces the overall pathfinding time (time taken to find paths between two nodes in the network) and routing time (time taken to route the payment). Due to space constraints we cover the usage of DHTs in file-sharing systems, MANETS (Mobile Ad-hoc Networks), and  VANETS (Vehicular Ad-hoc Networks) in the Appendix \ref{sec:appendrelated}.
 

All routing protocols have an overall routing complexity that is \emph{linear} in the number of nodes in the PCN in the worst case. Reducing this bound to \emph{sub-linear} while preserving the \emph{privacy of nodes and network topology} and also ensuring the \emph{atomicity} of the payments is a significant research challenge. 

Our contributions are: \\
\noindent
1) We design an efficient decentralized routing protocol, $\ripd$ with no trusted entities, using DHTs to reduce the routing time from $O(n)$ to $O (\log r + u)$, where $n$ is the total number of users/nodes in a PCN, $r$ is the total number of routing helpers (untrusted nodes that aid transaction routing), and $u$ is the number of non-routing helper nodes. \\
2) $\ripd$ preserves the privacy of nodes and their channel balances, as well as maintains privacy of the network topology.\\
3)  We experimentally demonstrate the scalability and efficiency of $\ripd$ using transaction data from the Ripple network~\cite{rippleapi}, and prove its security in the Universal Composability (UC) framework.

\textbf{Outline:}
In Section \ref{sec:related} we discuss relevant related work, in Sections \ref{sec:sysmodel} and \ref{sec:adversary}, we explain our system and adversary models respectively. In Sections \ref{sec:cons} and \ref{sec:proto} we present the construction of $\ripd$, in Section \ref{sec:impl}, we present our experimental evaluation. In Section \ref{sec:security} we analyze the security of $\ripd$ in the UC framework,
and in Section \ref{sec:conc}  we conclude the paper. 

%% file: related_full.tex
\section{Related Work}
\label{sec:related}

\begin{table*}[h!]
\caption{Comparison of Routing protocols in PCNs. TP: Topology privacy, ASR: Avoids source routing, DG: Disjoint graphs.}
\label{tbl:comaprison}
\centering
\begin{tabular}{|p{2.65cm}|p{2.00cm}|p{1.10cm}|p{1.20cm}|p{0.40cm} |p{0.60cm}|p{2.20cm}|p{1.50cm}|p{0.40cm}|}
\hline
Routing protocols              & Concurrency                 & Privacy & Balance security & TP        & ASR & Decentralized & Atomicity & DG \\ \hline
MPCN-RP~\cite{mpcnrp} &  $\xmark$ & $\xmark$ & $\checkmark$ & $\xmark$ & $\xmark$ & $\xmark$ & $\checkmark$ & $\xmark$ \\ \hline

Eckey \emph{et al.} ~\cite{Eckey2020SplittingPL} & $\xmark$                         & $\xmark$     & $\checkmark$      & $\checkmark$ & $\checkmark$             & $\checkmark$           & $\checkmark$   & $\xmark$    \\ \hline
Vein~\cite{vein}                          & $\xmark$                          & $\xmark$      & $\xmark$       & $\xmark$              & $\xmark$           & $\xmark$           & $\xmark$   & $\xmark$    \\ \hline
Auto tune~\cite{auto} &  $\xmark$ & $\xmark$ & $\xmark$ & $\xmark$ & $\xmark$ & $\xmark$ & $\xmark$ & $\xmark$ \\ \hline
Kadry 
~\emph{et al.} 
~\cite{machine}                    & $\xmark$                          & $\xmark$      & $\xmark$       & $\xmark$             & $\xmark$            & $\xmark$           & $\xmark$    &  $\xmark$  \\ \hline
FSTR~\cite{FSTR}                          & $\xmark$                          & $\xmark$      & $\xmark$       & $\xmark$              & $\xmark$             & $\xmark$           & $\xmark$  & $\xmark$       \\ \hline
SilentWhispers
~\cite{Malavolta2016SilentWhispersES}               & $\xmark$                          & $\checkmark$      & $\checkmark$      & $\checkmark$ & $\checkmark$             & $\checkmark$           & $\checkmark$  & $\xmark$       \\ \hline
Blanc~\cite{blanc}                         & $\checkmark$                         & $\checkmark$    & $\checkmark$      & $\checkmark$              & $\checkmark$             & $\checkmark$           & $\checkmark$  & $\xmark$     \\ \hline

SpeedyMurmurs ~\cite{roos2017settling}                & $\checkmark$                         & $\checkmark$     & $\xmark$      & $\checkmark$ & $\checkmark$             & $\checkmark$           & $\xmark$ & $\xmark$       \\ \hline
Spider~\cite{sivaraman2020high}                        & $\xmark$                          & $\xmark$      & $\xmark$       & $\xmark$              & $\xmark$             & $\xmark$           & $\xmark$ & $\xmark$      \\ \hline
Flash~\cite{flash} &  $\xmark$ & $\xmark$ & $\xmark$ & $\xmark$ & $\xmark$ & $\xmark$ & $\checkmark$ & $\xmark$ \\ \hline

Coinexpress
~\cite{coinexpress}                   & $\checkmark$ & $\xmark$      & $\xmark$       & $\checkmark$             & $\checkmark$             & $\checkmark$           & $\checkmark$  & $\xmark$       \\ \hline

Webflow~\cite{webflow}                       & $\xmark$                          & $\checkmark$     & $\xmark$      & $\checkmark$              & $\checkmark$             & $\checkmark$           & $\xmark$  & $\xmark$     \\ \hline

Robustpay~\cite{robustpay}                        & $\xmark$                          & $\xmark$      & $\checkmark$       & $\xmark$              & $\xmark$             & $\xmark$           & $\checkmark$ & $\xmark$      \\ \hline
Robustpay+~\cite{robustpay++}                        & $\xmark$                          & $\xmark$      & $\checkmark$       & $\xmark$              & $\xmark$             & $\xmark$           & $\checkmark$ & $\xmark$      \\ \hline

\textbf{\textit{RACED}}  \ & $\checkmark$ & $\checkmark$ &$\checkmark$ & $\checkmark$ & $\checkmark$ & $\checkmark$ &$\checkmark$ & $\checkmark$   
   \\ \hline

\end{tabular}
\end{table*}

\textbf{Routing protocols with security guarantees}: 
The security property that we want for routing protocols in PCNs is that honest parties should not lose funds because of malicious behavior by other parties in the system. To this end Malavolta \emph{et al.}~\cite{Malavolta2016SilentWhispersES} proposed a routing protocol leveraging trusted entities called \emph{landmarks}
to provide secure routing between the sender and the receiver. 
The landmark finds a path between itself and the sender and itself and the receiver; these sub-paths are combined to get the full path. 
The idea of using untrusted entities to facilitate routing has been proposed by Panwar \emph{et al.} in~\cite{blanc} that uses a set of well-connected nodes called \emph{routing helpers} to facilitate routing. However, this protocol  has 
a very high communication overhead during the pathfinding phase, in addition to using the blockchain as an auditing mechanism which makes it very expensive to deploy in the real-world.

Roos \emph{et al.} proposed a routing mechanism in \cite{roos2017settling} that uses graph embedding, where the routing is carried out by constructing a spanning tree of the entire network. While this work improved upon~\cite{Malavolta2016SilentWhispersES} by supporting concurrent transactions,
the sender picks a random amount to be transmitted along a path without knowing whether the path has sufficient liquidity, which could lead to a high rate of transaction failure.
Besides, frequently needing to update the embedding for a dynamic network topology results in a heavy computational overhead.
The routing protocol proposed by Pietrzak \emph{et al.} in \cite {lightpir} uses the idea of  Private Information Retrieval (PIR). The shortest paths between all the nodes are computed and stored in trusted servers which incur a large storage overhead. A honest majority is assumed among the servers. When a payment needs to be routed, the sender queries these trusted servers for  the available list of shortest paths to the intended receiver. This would also require the sender to download the complete network topology.
The protocol proposed by Subramanian \emph{et al.}~\cite{subramanian2020balance} leverages the idea of distributed hash tables to replenish the depleted link weights of nodes in a PCN, in a process called \emph{rebalancing}, and does not focus on pathfinding or routing of transactions, hence their work is orthogonal to $\ripd$. None of the aforementioned works can route transactions in disjoint graphs. 

\par
\noindent
\textbf{Routing protocols with no privacy/security guarantees}:
There are a few works that use breadth first search (BFS) or max-flow algorithms to design routing protocols for PCNs~\cite{FSTR,vein,coinexpress} but do not provide security/privacy of nodes in the PCN. Besides, using traditional max-flow algorithms such as Ford-Fulkerson (implemented using Edmonds-Karp method) and Goldberg-Tarjan algorithms incur significant overheads of $O(|V||E| ^{2})$~\cite{cormen2022introduction} and $O(|V|^{3})$~\cite{goldberg} respectively, in a graph $G(V,E)$, which is  not scalable to large PCNs. The ideas proposed by Abdelrahman \emph{et al.} in \cite{abdelrahman,aly2013securely,aly2015network} present distributed versions of Dijkstra's shortest path algorithm, and the minimum cost flow problem, both of which can be potentially applied to perform routing in PCNs. The distributed version of Dijkstra's shortest path algorithm has a computational complexity of $O(|V|^{2})+O(|V|)$, which makes it non-scalable to large scale PCNs. The computational complexity of the distributed version of the minimum cost-flow problem is $O(|V|^{8}log(|V|))$ and the communication complexity is $O(|V|^{10}log(|V|))$, which makes it infeasible to be applied for large scale PCNs. We give a detailed descriptions of the ideas proposed in  \cite{abdelrahman,aly2013securely,aly2015network} in Appendix \ref{sec:appendrelated}. 
The idea proposed in \cite{machine} by Kadry \emph{et al.} uses a machine learning-based approach to find a path between the sender and receiver. 
This work does not focus on route discovery but instead focuses on selecting the best path amongst the ones that have already been chosen using BFS. The work proposed in \cite{sivaraman2020high} uses buffers (called Spider Routers) in the form of queues to store and route transactions. However, it 
makes transactions wait for an indefinite amount of time before they are routed, besides it does not take into account the privacy of the nodes involved in a transaction. Other works such as~\cite{webflow} and ~\cite{Eckey2020SplittingPL} have been proposed that do not provide privacy of nodes in the PCN. 

RobustPay+~\cite{robustpay++} and its preliminary version, Robustpay~\cite{robustpay} focus on building a routing protocol for PCNs that supports multiple paths from a sender to a receiver from which the sender chooses only one path to route the payment. Their main contribution lies in constructing multiple paths such that there is no overlap in terms of nodes between any pair of paths. This is done to prevent transaction failures caused by nodes becoming unresponsive or going offline in the PCN.
The idea proposed by Chen \emph{et al.}, MPCN-RP \cite{mpcnrp}, focuses on building a source routing protocol that minimizes the transaction fees. This protocol presents a modified version of Dijkstra's algorithm, in which the length of the path (in terms of hop-count) is taken into consideration along with the edge weights. Unlike Robustpay and Robustpay+, \cite{mpcnrp} constructs only a single path from the sender to the receiver and the entire amount is routed along this path. Auto-Tune \cite{auto} proposed by Hong \emph{et al.} is a routing protocol that supports structured payments. Auto-Tune computes a total of $k$ shortest paths between a sender and receiver where $k$, an arbitrary number, is decided by the sender. Unlike \cite{mpcnrp}, the amount to be transacted is split across these multiple paths and is routed to the receiver. This work does not take into account the presence of redundant nodes along the $k$ shortest paths, which makes it different from the ideas in \cite{robustpay,robustpay++}.   We refer the reader to the Table \ref{tbl:comaprison} for the  differences between $\ripd$ and \cite{robustpay, robustpay++, mpcnrp, auto}.  

In Table \ref{tbl:comaprison}, we give a qualitative comparison between $\ripd$ and other routing protocols based on the comparison metrics defined as follows.
1) Concurrency: Concurrency is achieved when several transactions are routed simultaneously. 2) Privacy: Privacy is achieved when the identity of a node is not known to any other node in the network except its immediate neighbors. 3) Security: Security is achieved  when no honest party loses funds because of malicious behavior by other parties in the system. 4) Topology privacy (TP): Topology privacy is achieved when no node in the network knows the entire network topology. 5) Avoids source routing (ASR): Source routing is avoided when the sender does not construct the entire path from itself to the receiver. 6) Decentralization: Decentralization is achieved when there is/are no central entity/entities that construct the path for the sender. 7) Atomicity: Atomicity is achieved when all the link weights of the nodes along the transaction path go back to the state that they were in before the transaction was initiated in the event of a transaction failure. 8) Disjoint graph applicability (DG): A routing protocol is said to be applicable to disjoint graphs, if it works even when the network graph is not fully connected. For the routing protocols in Table \ref{tbl:comaprison}, we conjecture that support for concurrency, privacy and atomicity can be provided (in the protocols that do not already have them) by using HTLCs \cite{htlc} and the identity generation mechanism used in this paper. 
Modifying these protocols to achieve the remaining properties of topology privacy, avoiding source routing, decentralization, and making them applicable for disjoint graphs is non-trivial and is not a part of their design goals. 

%% file: sysmod_full.tex
\section{System Model}
\label{sec:sysmodel}
\begin{figure}[H]
 \centering
    \includegraphics[width=0.95\columnwidth]{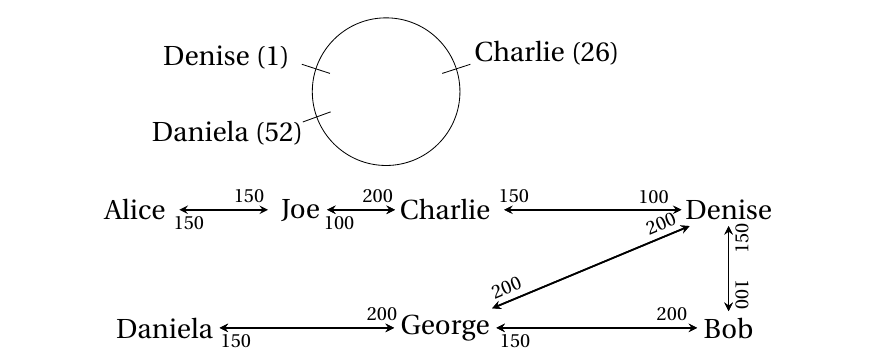}
    \caption{Three routing helpers in a DHT overlay over a PCN}
    \label{fig:fig1}
\end{figure}
In this section, we introduce the components of $\ripd$, the parties involved, and the terminology we use in the rest of the paper.
\par 
A PCN can be modeled as a directed graph where a directed edge from a node $i$ to $j$ with an edge weight of $\alpha$ signifies the balance of node $i$ in the payment channel between $i$ and $j$, denoted by $lw_{i,j}$ = $\alpha$. 
For instance, referencing Figure~\ref{fig:fig1},
in the link between Denise and Bob,
$lw_{\bob,\denise} = 150$ and 
$lw_{\denise,\bob} = 100$. 

\subsection{Parties}
\textbf{Routing Helpers ($\rh$)}: In $\ripd$,
a routing helper ($\rh$) is a  node that helps the sender and the receiver route transactions between each other. We define a dynamic set $\mathbb{RH}$ that contains all the routing helpers. $\rh$s in $\ripd$ are similar to the ``routing nodes" or trampoline nodes used by the real-world PCN, Lightning Network~\cite{pool}. If a node volunteers to become an $\rh$, it needs to join a Distributed Hash Table (DHT) overlay and establish channels with a few other nodes. This is independent of the underlying PCN topology. We do not assume $\rh$s are trusted, nor do we assume any honest majority among them. In $\ripd$, the $\rh$s charge a fees for providing their services and $\ripd$ is resilient to (n-2) $\rh$ failures for ``n" $\rh$s.
We organize all $\rh \in \mathbb{RH}$ as part of a DHT to ensure that they can route transactions in $O(\log \mathbb{|RH|})$ time, using consistent hashing to locate each other.
In this paper we instantiate the DHT using Chord~\cite{chord}, however, there are no technical impediments to using other DHT protocols such as Pastry \cite{rowstron2001pastry}, Kademlia \cite{maymounkov2002kademlia}, Tapestry \cite{zhao2004tapestry} and more. In Figure~\ref{fig:fig1}, we depict three routing helpers, Charlie, Denise and Daniela. The numbers adjacent to the routing helpers represent their unique identifiers inside the DHT ring.

\noindent 
\textbf{Sender and Receiver ($\send$, $\rec$)}: With respect to Figure \ref{fig:fig1}, the sender, Alice, is a node in the PCN who initiates a payment that needs to be routed across the network to receiver Bob. 
She  only knows the link weights of her immediate neighbors. 
Once a path has been found between Alice and Bob using $\ripd$, Bob generates parameters needed for establishing HTLCs (Hashed Time-Lock Contracts)~\cite{htlc} to complete the payment process. The purpose of establishing HTLC is to ensure atomicity of payments. We assume Alice and Bob can communicate with each other using a secure out-of-band communication channel, but can only do path-based routing of transactions. This is similar to real world PCNs, such as the Lightning Network~\cite{lightning}, where out-of-band communication channels are used by the receiver to communicate the digest required to complete the HTLC payment to the sender.

\noindent 
\textbf{End Routing Helper ($\erh$) and Nearest  Routing Helper ($\nearrh$)}:  $\erh$ is the routing helper from the DHT ring that is closest to Bob based on hop count. Similarly, $\nearrh$ is the nearest routing helper based on hop count to Alice. If we assume that the path taken is Alice $\rightarrow$ Charlie $\rightarrow \dots$ $\rightarrow$ Denise $\rightarrow$ Bob, the $\erh$ is Denise and $\nearrh$ is Charlie.



\noindent 
\textbf{Blockchain}: $\ripd$ can be deployed on any blockchain that is either permissioned or permissionless that supports HTLCs. $\ripd$ is compatible with the Lightning Network, which runs on top of the Bitcoin blockchain. In $\ripd$ we only use the blockchain for dispute resolution and it is not used during transaction routing and processing. \\ 
\subsection{Setup and Terminology}
\label{sec:setup}

\textbf{Keys setup}: In $\ripd$, every user $i$ in the PCN has a long-term signing and verification keypair denoted by ($\sk_{i}$,$\vk_{i}$), 
and a pseudonymous, temporary signing and verification keypair ($\SK_{i}$,$\VK_{i}$). In a decentralized network, each node generates its own keys. 
\footnote{For instance, in transactions involving Bitcoin in the Lightning Network, each node generates a long-term keypair on Bitcoin's secp256k1 elliptic curve \cite{Keypair}.} 
A node's long-term public key in $\ripd$ is used within the network to establish an encrypted and authenticated connection with its neighbors. The temporary keys in $\ripd$ provide pseudonymity and hide the real identity of the node from its non-neighboring nodes in the PCN.
To enable this, the temporary verification key is signed by the long-term signing key to produce a signature: $\sign_{\sk_{i}}$ ($\VK_{i}$) $\rightarrow$ $\sigma$. 
Each user $i$ exchanges its temporary and long-term verification key with all its neighbors, who verify $\sigma$ using $i$’s long-term verification key.
Two nodes that are not immediate neighbors, use their temporary signing keys to sign messages and their temporary verification keys to verify the corresponding signatures. If Alice intends to route a payment to Bob, we assume both of them will know each other's real identities, since a sender will not typically route a payment to an unknown receiver.

\noindent
\textbf{Immediate Neighbor}: Consider two nodes $i$ and $j$ that have a payment channel between them with the link weights denoted by $lw_{i,j}$ and $lw_{j,i}$. These two nodes are called immediate neighbors of each other. 

\noindent 
\textbf{Pathfinding and routing times}: We define the pathfinding time as the time taken to find a path involving several intermediate nodes between the $\send$ and the $\rec$. Routing time is defined as the time taken to route the payment after a path has been found. 

\noindent
\textbf{Routing fees}:  In PCNs, every node charges fees for forwarding the payment from its predecessor to its successor along the path; the fee structure varies according to the PCN being used. For instance, Lightning Network charges two types of fees, the base fee, which is fixed irrespective of the transaction amount, and rate fees that vary according to the amount being routed \cite{lightningfees}. In this paper, we assume a unit fee is charged per hop. This makes the routing fees equal to the path length.


%% file: advmod_full.tex
\section{Adversary Model}
\label{sec:adversary}

In this section, we outline the trust assumptions for the parties involved in $\ripd$, and state our security and privacy goals.
The sender and receiver in a transaction can be un-trusted and can arbitrarily deviate from protocol steps. Either of them can choose to abandon a transaction in-progress, or introduce delays in a transaction, with the goal of locking up collateral along paths.
In $\ripd$, we assume each sender and receiver have access to each other's real identities, and the receiver will know the amount, $\amount$ being transacted between them, since users do not send payments to unknown entities with unspecified amounts. All the nodes in the PCN, including the sender and the receiver, will know the real identities of all the routing helpers, $\rh$s in the DHT ring, and will also know the maximum amount that each $\rh$ can route to its finger table entries. Every node in the PCN, including the routing helpers will know the balances they have and will also know the balance of their immediate neighbor in the payment channel between the node and its immediate neighbor. In addition to this, the nodes in the PCN present along the path for routing a  transaction between a sender and a receiver will know the real identities of all their immediate neighbors and the amount being transacted between the sender and receiver along that path. The nodes in the PCN that are \emph{not} along the path for a transaction between the sender and receiver will only know the real identities of their immediate neighbors and will not have access to any  information regarding the transaction, such as the amount, transaction id, etc.

The $\rh$s in $\ripd$ can also be malicious. They can arbitrarily deviate from the protocols, although we assume at least two $\rh$s will be available at a given point of time to route transactions. For addressing distributed denial of service attacks where all the nodes in a DHT are taken down by an adversary, we refer the reader to existing mitigation strategies~\cite{Hakem,srivatsa,zied}. We also assume the adversary will be economically rational, i.e., it will always try to maximize its profit. The $\nearrh$ knows the pseudonymous identity of the sender and the $\erh$ will know the pseudonymous identity of the receiver. Other routing helpers (which are neither $\nearrh$ or $\erh$) will know the amount being routed for a transaction if they are present in the finger table of the $\nearrh$ or if they are present in the finger tables of $\rh$s which are present in the $\nearrh$'s finger table. If a $\rh$ is present in both the finger table of $\nearrh$ and $\erh$, it will have access to the amount being transacted.
We now give our security and privacy goals.

\noindent
\textbf{Defining our Security and Privacy Goals}.
 1) \textbf{Balance security}: No honest node along a transaction path should lose funds even if all the other nodes, including the intermediaries, and/or the sender, receiver, are malicious. If the $\nearrh$ or $\erh$ turn malicious at any point and decide to leak the identity of the sender or the receiver, respectively, it will only reveal their pseudonymous identities since the real identities of the sender and receiver are not known to any $\rh$s in the DHT ring. 
 2) \textbf{Sender/receiver privacy}: The real identities of the sender and the receiver are only known to each other and their immediate neighbors in the network.
 
 3) \textbf{Link privacy}: Every node only knows the balance in the channel it shares with its immediate neighbors. 4) \textbf{Atomicity}: If a transaction does not go through for any reason, all the link weights of the nodes along the transaction path should go back to the state that they were in before the transaction was initiated.


%% file: construction_full.tex


\section{Construction}
\label{sec:cons}

In this section, we present the challenges associated with leveraging DHTs for secure routing in PCNs and we describe the key ideas in $\ripd$ that solve these challenges and describe the detailed construction of $\ripd$.
 
To address the challenge identified in Section \ref{sec:intro}, our idea is to use a DHT comprising of $\rh$s which guarantees a logarithmic routing time. 
We note that it is non-trivial to apply DHTs to perform secure PCN routing due to the following challenges:

Challenge 1:  DHTs were designed to facilitate information sharing in a p2p network, whereas PCNs were developed for facilitating financial transactions between users. Nodes in the DHT communicate with each other using the standard IPv4 communication protocol. In PCNs, though nodes communicate with each other using the same standard, they also need to exchange payments between them for which the IPv4 communication standard cannot be used. As a solution to this challenge,  $\rh$s in $\ripd$  open a payment channel on the blockchain with each of the $\rh$s in their finger tables to facilitate payments.

Challenge 2: In DHTs, there is no notion of privacy; each peer in the DHT knows the details of the file (information) segments that every other peer is responsible for. Whereas in PCNs, the local channel balance of a node is known only to its immediate neighbor. In $\ripd$, to safeguard its local channel balance in a payment channel, each $\rh$ $i$ decides on a maximum amount that it can transact with its finger table entry $k$ and only this maximum amount is known to all the other nodes in the PCN, providing link privacy to the $\rh$s.

Challenge 3: In a DHT, the property of atomicity (defined in Section \ref{sec:adversary}) is not required since nodes only exchange information. Whereas in PCNs atomicity is very important since it ensures that no honest party loses their funds because of malicious behavior by other parties in the system. 
In $\ripd$ atomicity is ensured since $\rh$s (and all the other non-routing helper nodes) process payments between each other using HTLCs.

Challenge 4: Transacting money between nodes in the PCN causes a depletion in the balance of node, whereas, no such depletion exists in DHTs. We solve this challenge using the maximum amount computation described in Protocol \ref{algo:dhtsetup}.

Addressing these challenges and utilizing DHTs to reduce the overall routing complexity in PCNs requires careful design and is non-trivial.

We first give a brief overview that describes the working of $\ripd$ that solves all the aforementioned challenges at a high level and we follow it up with a detailed description of its construction. For the reader's easy reference, we give a table of notations in Table~\ref{tbl:notations}.
\begin{table}[H]
    \caption{Notations}
    \label{tbl:notations}
      \centering
        \begin{tabular}{|c|c|}
        \hline
        \textbf{Notation} & \textbf{Description} \\
        \hline\hline

        $\lambda$ & Security parameter \\
		\hline 
 	$\amount$ & Amount to be paid by the sender to the receiver\\
 	\hline  
 	$\mathbb{RH}$ & Set of routing helpers \\
 	\hline
 	$n$ &  Number of nodes in the PCN\\ 
 	\hline 
  $\mathbb{I}_{i}$ & Set of immediate neighbors of a node $i$ in the PCN \\
  \hline
 
 	$(\SK_{i}$,$\VK_{i})$, $(\sk_{i}$, $\vk_{i})$ & Temporary and long-term signing/verification keypair of node i \\
 	\hline

 		$\erh$, $\nearrh$ & End routing helper and nearest routing helper, respectively \\
 		\hline

 
 		$\delta$ \ & Time interval for signature generation  \\
 		\hline 
 		$\maxamount_{i,j}$ \ & Maximum amount that can be transacted between nodes $i$ and $j$ \\
 		\hline 

 		$\hc_{i,j}$ & Number of hops between nodes $i$ and $j$ \\
 		\hline 
 		$\txid$ & Transaction identifier \\
 		\hline 
 		 
		$tc_{i,j}$ & Current timestamp for signature created on $\maxamount_{i,j}$\\
		\hline 
		$tv_{i,j}$ &  Time until which the signature created on $\maxamount_{i,j}$ is valid\\
  \hline
  
        \end{tabular}
    
\end{table}

In $\ripd$, we instantiate the DHT using Chord \cite{chord}. Due to space constraints, we give an overview of Chord in the Appendix \ref{sec:appendchord}.

\subsection{Technical Overview}
\begin{figure}[H]
\centering
\includegraphics[width=0.75\columnwidth]{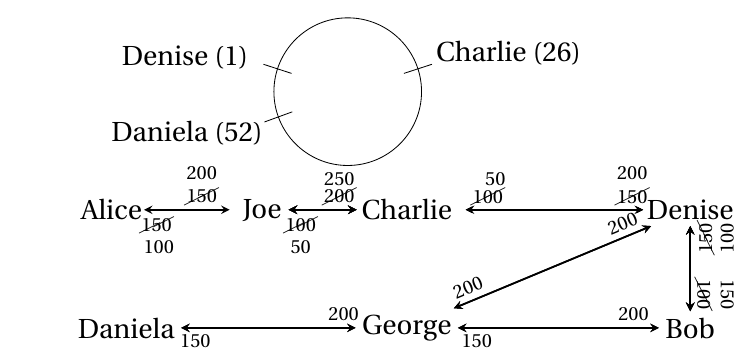}
    \caption{Alice transmitting 50 coins to Bob via $\rh$s Charlie and Denise in the DHT ring.}
    \label{fig:fig2}
\end{figure}

Let us consider a sender Alice in a PCN, as depicted in Figure~\ref{fig:fig2}, who intends to route an amount, $\amount = 50$ coins to a receiver, Bob. We use $\rh$s, Charlie, Daniela, and Denise that belong to a set $\mathbb{RH}$, to route the payment. During the key-generation and setup phase (described in Protocol \ref{algo:keysetup}), each node creates a pair of long-term and temporary signing and verification keys. The temporary identity of a node is tied into the long-term identity as described in Section \ref{sec:sysmodel}. This is done to hide the real identity of a node in the PCN from its non-neighboring nodes, which helps in achieving the goal of sender/receiver privacy as described in Section \ref{sec:adversary}.
In the DHT setup phase (described in Protocol \ref{algo:dhtsetup}), the first node that volunteers to be an $\rh$ establishes the DHT overlay by creating a unique identifier (depicted as the number next to an $\rh$'s name) and populating its local hash table. This local hash table is termed as a node's \emph{finger table} in Chord~\cite{chord}. 
Since we use Chord to instantiate the DHT in $\ripd$, for terminological consistency, we refer to a node's local hash table as finger table in the rest of the paper (we give a detailed description of Chord in Appendix \ref{sec:chord}). All the subsequent nodes that volunteer to be $\rh$s join the DHT, create unique identifiers and populate their finger tables with the identifiers of other $\rh$s. The $\rh$s create signatures (using their long-term signing keys) on the maximum amount that they are willing to route to the other $\rh$s in their finger tables. This is done to hide the actual channel capacities between the $\rh$s in the DHT, which helps us in achieving our goal of link privacy.

Alice finds a path to the $\rh$ nearest to her, $\nearrh$ using any of the existing constructions such as \cite{blanc,roos2017settling,Malavolta2016SilentWhispersES}. These existing routing algorithms have an end-to-end worst-case pathfinding time complexity of $O(n)$, where $n$ is the number of nodes in the PCN. Our goal in this paper is to improve this worst-case upper bound by using DHTs. Inside the DHT, the worst case pathfinding complexity is logarithmic in the number of nodes, $O(\log |\mathbb{RH}|)$ which
 improves the overall complexity of pathfinding. Hence, we focus on routing inside the DHT ring, and assume the sender/receiver can find a path to $\rh$s using existing methods. In Figure \ref{fig:fig2} we assume the $\nearrh$ is Charlie. Alice requests Charlie to find paths from himself to all the other $\rh$s in the DHT ring, i.e., Denise and Daniela. Charlie has to find paths such that the amount, $\amount$ to be sent by Alice is less than or equal to the maximum amount, $\maxamount_{i,k}$ that each $\rh$ $i$ on a given path is willing to route to the next $\rh$ $k$ in the path. Charlie finds the paths in two phases. In Phase 1, Charlie finds paths from himself to all the $\rh$s that are a part of his finger table and that can route the amount, $\amount$ specified by Alice and adds these $\rh$s to a stack $\pset$ maintained locally by him. In Phase 2, Charlie finds paths from himself to the $\rh$s that are \emph{not} part of his finger table but can still route the amount, $\amount$ specified by Alice.
 These two phases are described in detail in Protocol \ref{algo:path1}. Once the pathfinding phases are complete, the stack $\pset$ that contains the list of paths and signatures is sent to $\alice$. Alice then verifies the signatures of the $\rh$s in the stack $\pset$ on the maximum amount that they can route to the $\rh$s in their finger tables. 

Upon successfully verifying the signatures, Alice sends the $\erh$ in each path to Bob via a secure out-of-band communication channel. Bob then picks the $\erh$ that is nearest to him based on hop count. In Figure \ref{fig:fig2}, the $\erh$ is Denise. Bob notifies Alice about his choice of the $\erh$ via a secure out-of-band communication channel. Alice then picks the path containing the $\rh$ picked by Bob as $\erh$.
This finalizes the path along which $\amount$ needs to be routed inside the DHT ring. At any point, if the signatures do not verify, Alice writes the publicly verifiable signature details to the blockchain, at which point they can be verified by miners, and others involved in the system. This ensures that cheating $\rh$s will be caught, and made to leave the DHT ring. Once Alice and Bob have agreed on the $\erh$ Denise, Bob chooses a random preimage $X$, and hashes it to produce a digest $Y$. The payment mechanism is initiated using HTLCs. Using HTLCs ensures that no honest party loses any funds because of malicious behavior by other parties in the system, which helps us in achieving our  goal of balance security. HTLCs also ensure that all the link weights of the nodes along the transaction path go back to the state they were in prior to the commencement of the transaction if the transaction fails for any reason. This achieves our goal of atomicity. 

\subsection{Helper functions}

We now describe the helper functions used in the protocols of $\ripd$.

\noindent
1) \underline{{$\mathsf{ChoosePath}$ ($\mathbb{P}$, $\erh$) $\rightarrow$ $\mathbb{P ^ \prime}$}}: This function picks a path for routing the payment. It takes two parameters, stack $\pset$, and the $\erh$ as inputs, and returns a path $\mathbb{P}^\prime$ that contains the $\erh$ as the last node in the path. If multiple paths with the same $\rh$ as the $\erh$ are present, the last $\rh$, it returns the shortest path. \\
2) \underline{{ $\mathsf{NH}$ ($i, j$)$\rightarrow$ $hc_{i,j}$}}: This function calculates the hop count between two nodes, $i$ and $j$ inside the DHT ring. \\    
3) \underline{{$\open$ ($\VK_{i}$, $\VK_{j}$, $lw_{i,j}$, $lw_{j,i}$) $\rightarrow$ \{$success$, $failure$\}}}: This is a function that opens a new payment channel between two nodes. It takes in the temporary verification keys of the nodes involved in opening the channel, denoted by $\VK_{i}$ and $\VK_{j}$, and the amounts being deposited on the links as input parameters. The nodes interested in opening a payment channel create a transaction tuple that contains the $\VK$ of the nodes and the amounts they individually intend to deposit in the channel. This tuple is signed by both nodes with their temporary signing keys, $\SK_{i}$ and $\SK_{j}$, and is posted to the blockchain. The two nodes involved in the opening of the payment channel sign a single transaction tuple, making it a 2-2 multisig transaction. For representational clarity, we have abstracted the description of blockchain writes. Upon a successful opening of a payment channel between the  nodes, this function returns a $succcess$. \\
4) \underline{ $\htlcpay$ ($\vk_{i}$, $\vk_{j}$, $\txid$, $\amount$) $\rightarrow$ \{$success$, $failure$\}}: This function completes the payment between two nodes once the preimage used to create the digest is revealed by one node to another. It takes the long-term verification key of the payer, $\vk_{i}$, the long-term verification key of the payee, $\vk_{j}$, the unique transaction id $\txid$, and the $\amount$ being transacted as inputs. Once the payee has revealed the preimage using which the digest was produced, this function checks if the preimage being revealed is correct. If yes, the payer updates the link weights between them accordingly. Upon a successful release of the correct preimage by the payee and the updating of the link weights by the payer, it returns $success$. Else it returns a $failure$. \\
5) \underline{$\lookup$ (${i}$, ${j}$) $\rightarrow$ $l$}: This function performs the lookup operation for a node nearest to a destination node based on node identifier. It takes the node identifier of the source denoted by ${i}$ and the node identifier of the destination denoted by ${j}$ as the inputs, and returns the node $l$ who has the largest node identifier and is less than or equal to ${j}$ from the finger table of ${i}$. \\
6) \underline{$\retrieve$ ($i$) $\rightarrow$ $\mathcal{B}_{i}$}: This function retrieves the entries of a node's finger table. This function takes node identifier $i$ of the node in the DHT ring as an input and returns a stack containing the node identifiers of the entries present in $i$'s finger table. \\ 
7) \underline{$\retrievenext$ ($\mathcal{B}$) $\rightarrow$ $i$}: This function takes a list as an input argument and returns the node identifier of the element that the head of the list points to.  \\
8) \underline{$\retrieveneighbors$ ($\vk_{i}$) $\rightarrow$ $\mathbb{I}_{i}$}: This function is used to retrieve the immediate neighbors of a node. It takes the long-term verification key of a node $i$, $\vk_{i}$, as an input and outputs a list, $\mathbb{I}_{i}$ containing the verification keys of the immediate neighbors of the node $i$. \\
9) \underline{$\succlookup$ (${i}$) $\rightarrow$ ${j}$}: This function looks up the successor of a node in the DHT ring. It takes the node identifier of a node denoted by ${i}$, and returns its immediate successor which is defined as the smallest node identifier in the DHT ring that is larger than ${i}$. \\ 
10) \underline{$\search$ (${i}$,${j}$) $\rightarrow$ \{$success$, \ $failure$\}}: This function searches for the presence of a node in another node's finger table. It takes in the node identifier of the node calling this function $i$ and the node identifier of the node being searched $j$ as inputs, and returns $success$ if $j$ is present in the finger table of $i$.

%% file: algo_full.tex
\section{Protocols}
\label{sec:proto}


 







$\ripd$ consists of seven protocols: $\mathsf{Key \ Setup}$ (Protocol \ref{algo:keysetup}), $\mathsf{DHT \ Setup}$ (Protocol \ref{algo:dhtsetup}), $\mathsf{DHT \ Processing}$ (Protocol \ref{algo:proofcompute}), $\mathsf{Find \ Path}$ (Protocol \ref{algo:path1}), $\mathsf{Path \ Validation}$ (Protocol \ref{algo:proof verify}), $\mathsf{Node \ Joining \ And}$  $\mathsf{ Node \ Leaving}$ (Protocol \ref{algo:nodejoin}), $\mathsf{Routing \ Payment}$ (Protocol \ref{algo:routepay}).  \\

The protocol $\mathsf{Key \ Setup}$ handles the generation of long-term and pseudonymous identities for all the nodes in the PCN. These keys are used to sign and verify messages in the subsequent protocols of $\ripd$. The steps of this protocol are self-explanatory and we give the protocol and its full description in Appendix \ref{sec:appendixproto}.

\noindent
\underline{$\mathsf{DHT \ Setup}$, Protocol \ref{algo:dhtsetup}}: This protocol handles the DHT ring setup and the computation of signatures on the maximum amount each $\rh$ can route to the $\rh$s in its finger table.
The DHT ring setup facilitates the joining of nodes as $\rh$s and the signatures created on the maximum amount from this protocol are used by the $\send$ while selecting a suitable path to route the $\amount$ to the $\rec$. Each node $i$ in the PCN that volunteers to be an $\rh$ hashes its IP address, $ipaddress_{i}$ with a collision-resistant and consistent hash function. The resulting digest of the hash becomes the node identifier of the $\rh$. Each $\rh$ $i$ locally maintains two stacks $\mathcal{B}_{i}$ and $\mathcal{J}_{i}$, and a list $\mathcal{L}_{i}$. $\rh$ $i$ will then compute the list of $\rh$s that are a part of its finger table and adds them to the stack $\mathcal{B}_{i}$, which contains repeated entries. The unique entries from $\mathcal{B}_{i}$ are added to the stack $\mathcal{J}_{i}$. $\rh$ $i$ creates signatures on the $\maxamount$ amount that it can route to the $\rh$s in the stack $\mathcal{J}_{i}$. The computation of the maximum amount and the corresponding signatures is done to hide the actual channel capacities between the nodes in the DHT, which helps us in achieving our goal of \emph{link privacy}. These signatures are created by $i$ and the $\rh$s in its finger table using their long-term signing keys and are added to the list $\mathcal{L}_{i}$. In addition, two timestamps, $tc$, which is the timestamp at which the signature was created, and $tv$, which is the timestamp until which the signature is valid, are added to $\mathcal{L}_{i}$.  
This protocol can be run in parallel by all the $\rh$s in the DHT ring. \\
 \begin{algorithm}[htbp]
\caption{$\mathsf{DHT \ Setup}$}
\label{algo:dhtsetup}
\DontPrintSemicolon
 \SetAlgoLined

All $\rh$s decide on the value of $\delta \sample \mathbb{R}^+$ and a hash function $H$: $\{0,1\}^\lambda$ $\rightarrow$ $\{0,1\}^m$\\ 
\For{i = 1;i  $\leq$ $|\mathbb{RH}|$;i++}
{
node $i$ that joins the DHT ring hashes its $ipaddress_{i}$ and creates a digest $Y_{i}$: $H(ipaddress_i$)$\rightarrow$ $Y_{i}$ \\
node identifier of $i$ = $Y_{i}$ \\
node $i$ will broadcast $Y_{i}$ and $\vk_{i}$ to all the nodes in the PCN. \\
 node $i$ maintains $\mathcal{B}_{i}$ = $\emptyset$, $\mathcal{J}_{i}$ = $\emptyset$ and $\mathcal{L}_{i}$ = $\emptyset$ \\
 $\forall\, j \in [1..m]$ node $i$ does $\mathcal{B}_{i}$.$\add$(${i}$ + $2 ^ {(j-1)}$ $\mod$ $m$)  \\
node $i$ does $\removeduplicate$($\mathcal{B}_{i}$)$\rightarrow$ $\mathcal{J}_{i}$ \\
 \While{($\mathcal{J}_{i}$.$\mathsf{empty}$ = False)}
 {

$\pop$.($\mathcal{J}_{i}$) $\rightarrow$ $k$\\

node $k$ does $\sign_{\sk_{k}}$($\maxamount_{i,k}$) $\rightarrow$ $\sigma_{\maxamount_{i,k}}^{k}$  \\ 

node $i$ does $\sign_{\sk_{i}}$($\maxamount_{i,k}$) $\rightarrow$ $\sigma_{\maxamount_{i,k}}^{i}$ and
 
node $i$ does
$\mathcal{L}_{i}$.$\add$($\maxamount_{i,k}$, $\sigma_{\maxamount_{i,k}}^{i}$, $\sigma_{\maxamount_{i,k}}^{k}$, $tc_{i,k}$, $tv_{i,k}$) \\

 }

}

\end{algorithm}
\begin{algorithm}[H]
\caption{$\mathsf{DHT \ Processing}$}
\label{algo:proofcompute}
\DontPrintSemicolon
\SetAlgoLined
\SetKwInOut{Input}{Input}
\SetKwInOut{Output}{Output}
\SetKwInOut{Parties}{Parties}




\tcc{All members of $\rh$ check this condition}


\For{i=1;i $\leq$ $|\mathbb{RH}|$;i++}
{
 node $i$ does $\retrieve$(${i}$)$\rightarrow$ $\mathcal{B}_{i}$ and 
$\removeduplicate$($\mathcal{B}_{i}$) $\rightarrow$ $\mathcal{J}_{i}$ \\

\While{$\mathcal{J}_{i}$.$\mathsf{empty}$ = False}
{
node $i$ retrieves ($\maxamount_{i,k}$, $\cdot$, $\cdot$, $\cdot$, $\cdot$) from $\mathcal{L}_{i}$\\
\If{$\maxamount_{i,k}$ $>$ $lw_{i,k}$}
{

 node $i$ assigns the maximum amount that can be routed to node $k$  to $\maxamount_{i,k} ^ \prime$ and does 
 $\sign_{\sk_{i}}$($\maxamount_{i,k} ^ \prime $) $\rightarrow$ $\sigma_{\maxamount_{i,k} ^ \prime}^{i}$ \\ 
node $k$ does
$\sign_{\sk_{k}}$($\maxamount_{i,k} ^ \prime $) $\rightarrow$ $\sigma_{\maxamount_{i,k} ^ \prime}^{k}$ and 
 node $i$ does
$\lset_{i}$.$\delete$($\maxamount_{i,k}$, $\sigma_{\maxamount_{i,j}}^{i}$, $\sigma_{\maxamount_{i,k}}^{k}$, $tc_{i,k}$, $tv_{i,k}$) and
$\lset_{i}$.$\add$($\maxamount_{i,k} ^ \prime$, $\sigma_{\maxamount_{i,k} ^ \prime}^{i}$, $\sigma_{\maxamount_{i,k} ^ \prime}^{k}$, $tc_{i,k} ^ \prime$, $tv_{i,k} ^ \prime$) \\

}

}

}

\If{($curr_{time}$ $\mod$ $\delta$ = 0) }
{


{

\For{i=1;i $\leq$ $|\mathbb{RH}|$;i++}
{

 node $i$ does $\retrieve$(${i}$)$\rightarrow$ $\mathcal{B}_{i}$ and 
$\removeduplicate$($\mathcal{B}_{i}$) $\rightarrow$ $\mathcal{J}_{i}$ \\

\While{($\mathcal{J}_{i}$.$\mathsf{empty}$ = False)}
{
node $i$ does $\pop$.$\mathcal{J}_{i}$ $\rightarrow$ $k$ \\
node $i$ retrieves ($\maxamount_{i,k}$, $\cdot$, $\cdot$, $\cdot$, $\cdot$) from $\mathcal{L}_{i}$\\
 node $i$ assigns the maximum amount that can be routed to node $k$ for the current time epoch to $\maxamount_{i,k} ^ \prime$\\

\If{($\maxamount_{i,k}$ == $\maxamount_{i,k}^\prime $)}
{

$tv_{i,k} ^ \prime$ = $tv_{i,k}$+ $\delta$ \\
Update ($\maxamount_{i,k}$, $\cdot$, $\cdot$, $\cdot$, $tv_{i,k}$) with ($\maxamount_{i,k}$, $\cdot$, $\cdot$,  $\cdot$, $tv_{i,k} ^ \prime$) in $\lset_i$

}

\Else
{

 node $i$ does  
$\sign_{\sk_{i}}$($\maxamount_{i,k} ^ \prime $) $\rightarrow$ $\sigma_{\maxamount_{i,k} ^ \prime}^{i}$  \\ 
node $k$ does
$\sign_{\sk_{k}}$($\maxamount_{i,k} ^ \prime $) $\rightarrow$ $\sigma_{\maxamount_{i,k} ^ \prime}^{k}$\\ 
 node $i$ does
$\lset_{i}$.$\delete$($\maxamount_{i,k}$, $\sigma_{\maxamount_{i,k}}^{i}$, $\sigma_{\maxamount_{i,k}}^{k}$, $tc_{i,k}$, $tv_{i,k}$) and
$\lset_{i}$.$\add$($\maxamount_{i,k} ^ \prime$, $\sigma_{\maxamount_{i,k} ^ \prime}^{i}$, $\sigma_{\maxamount_{i,k} ^ \prime}^{k}$, $tc_{i,k} ^ \prime$, $tv_{i,k} ^ \prime$) \\

return $\lset_{i}$ \\

}

}
}
}
}
\Else
{

do nothing
}
\end{algorithm}
\underline{$\mathsf{DHT \ Processing}$, Protocol \ref{algo:proofcompute}}: This protocol handles the creation of signatures on the new value of maximum amount, $\maxamount ^ \prime_{i,k}$, that each $\rh$ $i$ can route to each $\rh$ $k$ in its finger table once the time epoch $\delta$ expires, or when the maximum amount, $\maxamount_{i,k}$ that can be routed from a $\rh$ $i$ to its finger table entry $k$ exceeds the link weight (balance) that an $\rh$ has in the payment channel with its finger table entries. The signatures created in this protocol will be used by the $\send$ to check that the liquidity that exists between $\rh$s is sufficient to route the $\amount$ specified by her. 
The value of $\delta$ is a system parameter that the $\rh$s decide during the $\mathsf{DHT \ Setup}$ phase. Once the time epoch expires or when the maximum amount, $\maxamount_{i,k}$ that can be routed between an $\rh$ $i$ and its finger table entry $k$ exceeds the balance, $lw_{i,k}$ that the $\rh$ $i$ has in the payment channel between $i$ and $k$, each $\rh$ $i$ (that satisfies either of the two conditions in lines 5 or 15,) in the DHT ring retrieves the $\rh$s from its finger table, removes the duplicate entries, and adds the unique ones to the stack $\mathcal {J}_{i}$. Each $\rh$ $i$ will compute the signatures on $\maxamount ^ \prime_{i,k}$ for each $\rh$ present in its finger table. These signatures are computed by the $\rh$s using their long-term signing keys. After the computation of signatures, the signatures attesting to the $\maxamount$ in $\mathcal{L}_{i}$ from the previous time epoch are replaced with the new ones. In addition, the previous time stamps $tc$ and $tv$ are replaced with the fresh ones in the list $\mathcal{L}_{k}$. If for any reason, the $\maxamount_{i,k}$ between a $\rh$ $i$ and its finger table entry $k$ has not changed from the previous time epoch, only the time stamp of the signature validity, $tv$ will be incremented by $\delta$ and updated in the list $\mathcal{L}_{i}$. This protocol can be run by each $\rh$ in parallel. \\
\begin{algorithm}[htbp]
\caption{$\mathsf{Find \ Path}$}
\label{algo:path1}
\DontPrintSemicolon
\SetAlgoLined


Alice does $k \sample \{0,1\}^{\lambda}$, $H(k)$ $\rightarrow$ $\txid$ and sends ($\txid$, $\amount$, $\sigma_{\amount}$, $\VKalice$) to $\nearrh$, initializes a list $\mathbb{K}$ = $\emptyset$, creates a tuple ($\vk_{\rec}$, $\txid$, $\amount$) and adds it to $\mathbb{K}$  \\
$\nearrh$ initializes  stacks $\pset$, $\mathbb{W}$ = $\emptyset$ and does
\If{($\verify_{\VK_{Alice}}$($\amount$, $\sigma_{\amount}$) $\rightarrow$ 0)}
{
return $\bot$
}
\Else{
 $\retrieve$($\nearrh$) $\rightarrow$ $\mathcal{B}_{\nearrh}$\\
 $\forall$ $\rh$ $\curi$ $\in$ $\bset_{\nearrh}$, $\nearrh$ sends ($\txid$, $\amount$, $\sigma_{\amount}$, $\VKalice$) to each node $i$ and 
$\forall$ $j$ $\in$ $\mathbb{RH}$ and $j \notin$ $\mathcal{B}_{\nearrh}$ does $\mathbb{W}$.$\push$($j$)\\

 
\While{($\mathcal{B}_{\nearrh}$.$\mathsf{empty}$ = False) }
{
each node $i$ $\in$ $\mathcal{B}_{\nearrh}$ does \If{($\verify_{\VK_{Alice}}$($\amount$, $\sigma_{\amount}$) $\rightarrow$ 0)}
{
return $\bot$
}
\Else
{
$\pop$.($\mathcal{B}_{\nearrh}$) $\rightarrow$ $\curi$ \\
\If{($\amount$ $\leq$ $\maxamount_{\nearrh, \curi}$)}
{

$\nearrh$ picks $\pid$ $\sample$ $\{0,1\}^\lambda$ and does 
$\pset$.$\push$($\txid$, $\pid$, $\nearrh, \curi$, $\maxamount_{\nearrh, \curi}$, $\sigma_{\maxamount_{\nearrh, \curi}} ^ \curi $, $\sigma_{\maxamount_{\nearrh, \curi}} ^ \nearrh $, $tc_{\nearrh, \curi}$, $tv_{\nearrh, \curi}$, $\VKalice$ )\\
}

}

}
\While{($\mathbb{W}$.$\mathsf{empty}$ = False)}
{

$\nearrh$ checks  \If{($\mathrm{FindPathResp}$, $\curk$, $\txid$) has been received}
{
    $\nearrh$ does $\curi$ = $\curk$, sends ($\txid$, $\mathrm{FindPathReq}$, $\curp$) to $\curi$ \\
    }
    \Else
{

$\nearrh$ does $\pop$.($\wset$)$\rightarrow$  $\curp$, $\lookup$($\nearrh$,$\curp$) $\rightarrow$ $\curi$
 }
 $\nearrh$ checks 
\If{($\amount$ $\leq$ $\maxamount_{\nearrh,\curi}$)}
{
$\nearrh$ sends ($\txid$, $\mathrm{FindpathReq}$, $\curp$) to $\curi$ \\
\tcc{Node $\curi$ runs steps 21-25}
}
\If{($\mathsf{NH}$($\curi$,$\curp$ == 1))}
    {

 \If{($\amount$ $\leq$ $\maxamount_{\curi,\curp}$)}
        {  
            {
$\curi$ retrieves $(\maxamount_{\curi,\curp}$, $\sigma_{\maxamount_{\curi, \curp}} ^ \curi$, $\sigma_{\maxamount_{\curi, \curp}} ^ \curp$, $tc_{\curi,\curp}$ and $tv_{\curi,\curp})$ from $\mathcal{L}_{i}$ and constructs a tuple $\mathbb{Q}_{\curi}$ = ($\curi$, $\curp$, $\maxamount_{\curi,\curp}$, $\sigma_{\maxamount_{\curi,\curp}^\curi}$, $\sigma_{\maxamount_{\curi,\curp}^\curp}$, $tc_{\curi,\curp}$, $tv_{\curi,\curp}$) and sends the tuple ($\txid$, $\mathrm{FindpathResp}$, $\mathbb{Q}_{\curi}$) to $\nearrh$. $\nearrh$ samples $\pid$ $\sample$ $\{0,1\}^\lambda$ and does $\pset$.$\push$($\txid$, $\pid$, $\mathbb{Q}_{\curi}$) \\

            }
        }
        \Else
        {

        $\curi$ construct a tuple ($\txid$, $\mathrm{FindpathResp}$, $\curi$, $\curp$, $\bot$) sends it to $\nearrh$\\
        }
    }
\Else{
 $\curi$ does $\lookup$($\bset_{\curi}$) $\rightarrow$ $\curk$, sends ($\txid$, $\amount$, $\VKalice$) to $\curk$ 
 \If{($\amount$ $\leq$ $\maxamount_{\curi,\curk}$)}
   {
    $\curi$ constructs a tuple $\mathbb{Q}_{\curi}$ = ($\curi$, $\curk$, $\maxamount_{\curi,\curk}$, $\sigma_{\maxamount_{\curi,\curk}}^ \curi $, $\sigma_{\maxamount_{\curi,\curk}}^ \curk $, $tc_{\curi,\curk}$, $tv_{\curi,\curk}$,$\VKalice$) and sends ($\txid$, $\mathsf{FindpathResp}$, $\curk$, $\mathbb{Q}_{\curi}$) to $\nearrh$

     }

}
}
}
$\nearrh$ sends $\mathbb{P}$ to Alice\\

\end{algorithm}
\underline{$\mathsf{Find \ Path}$, Protocol \ref{algo:path1}}:
This protocol finds paths from the $\nearrh$ to all the other $\rh$s in the DHT ring that can route the $\amount$ specified by the $\send$. Initially, the $\send$, Alice, creates a random transaction id, $\txid$, contacts the $\nearrh$ and sends the $\txid$, the $\amount$ that she intends to route, her signature, $\sigma_{\amount}$, and her temporary verification key $\VKalice$ through a path-based transaction to $\nearrh$. The $\vk_{\rec}$ (in this case Bob), $\txid$, $\amount$ are added to a list $\mathbb{K}$ by Alice. Alice can be involved in several transactions with several receivers. The list $\mathbb{K}$ helps Alice maintain a record of all the transactions in which she is the sender. The path from the $\send$ to the $\nearrh$ can be calculated using the constructions described in \cite{blanc,Malavolta2016SilentWhispersES, roos2017settling}; we do not describe this process in this paper. The $\nearrh$ locally maintains two stacks $\mathbb{P}$ and $\mathbb{W}$. Upon successful verification of the signature of Alice on the $\amount$, the $\txid$, $\amount$ is sent by $\nearrh$ to all the $\rh$s in its finger table. For each entry, in its finger table, the $\nearrh$ checks if $\amount$ is less than the $\maxamount$ that the $\nearrh$ can route to that $\rh$ in $\mathcal{B}_{\nearrh}$ (stack containing $\nearrh$'s finger table entries). For each such entry, the $\nearrh$ randomly samples a path identifier, denoted by $\pid$ and adds the $\pid$, the node identifier of $\nearrh$, node identifier of the $\rh$s in its finger table, the corresponding $\maxamount$, the signatures of the $\nearrh$ and the $\rh$ on the $\maxamount$, the time stamps of signature creation and signature validity and the tuple is pushed on to the stack $\mathbb{P}$. In this manner, $\nearrh$ finds paths from to all $\rh$s in its finger table. $\nearrh$ now finds paths from itself to the $\rh$s in the stack $\mathbb{W}$. The $\nearrh$ pops the first $\rh$ in $\mathbb{W}$, denoted by $\curp$, and selects the $\rh$ closest to $\curp$ based on its node identifier and assigns it to $\curi$. The $\nearrh$ checks if the liquidity between $\nearrh$ and $\curi$ is suitable to route the $\amount$ specified by Alice and sends a tuple consisting of the $\txid$, $\mathrm{FindpathReq}$ message and the node identifier of $\curp$ to $\curi$. This transfers the control flow from $\nearrh$ to $\curi$. $\curi$ then checks for the presence of $\curp$ in its finger table. If present, $\curi$ constructs a tuple $\mathbb{Q}_{\curi}$ = ($\curi$, $\curp$, $\maxamount_{\curi,\curp}$, $\sigma_{\maxamount_{\curi, \curp}} ^ \curi$, $\sigma_{\maxamount_{\curi, \curp}} ^ \curp$, $tc_{\curi, \curp}$, $tv_{\curi,\curp}$) and sends this tuple to $\nearrh$ along with $\mathrm{FindpathResp}$ response message. If the $\rh$ $\curp$ is not present in the finger table of $\curi$, $\curi$ looks up the closest $\rh$ to $\curp$ in its finger table based on node identifier. This $\rh$ is denoted by $\curk$. $\curi$ checks if the liquidity between $\curi$ and $\curk$ is sufficient to route the $\amount$ specified by Alice, if yes, $\curk$ is sent to $\nearrh$. The $\nearrh$ then assigns the node identifier of $\curk$ to $\curi$. This process is repeated until a suitable path to $\curp$ is found. If a path to $\curp$ is found, the $\nearrh$ samples a path identifier, $\pid$, at random. The $\txid$, $\pid$ and all the tuples $\mathbb{Q}_{\curi}$, for each $\rh$ $\curi$ generated until this point are pushed on to the stack $\mathbb{P}$. Phase 2 is repeated until the stack $\mathbb{W}$ becomes empty, upon which $\nearrh$ sends $\mathbb{P}$ to Alice. The $\nearrh$ can turn malicious at any point in any of these phases and try to manipulate the contents of any tuple returned by the $\rh$s. However, the malicious behavior of $\nearrh$ will be caught when Alice verifies the amounts and signatures in Protocol \ref{algo:proof verify}. \\
 
\begin{algorithm}[htbp]
\caption{$\mathsf{Path \ Validation}$}
\label{algo:proof verify}
\DontPrintSemicolon
\SetAlgoLined

Alice maintains a list $\mathbb{W}_{Alice}$ = $\emptyset$ \\


Alice receives $\pset$ from the $\nearrh$\\

\While{($\mathbb{P}$.$\mathsf{empty}$ = False)}
{

Alice does $\pop$.($\pset$) $\rightarrow$ $\txid$ \\
Alice does \If{$\pop$.($\pset$)$\rightarrow$ $\pid$ }
{

Alice records a message ($\mathsf{New~Path}$) and does $\mathbb{W}$.$\add$($\pid$) \\
}

\Else
{

Alice maintains a list $\mathbb{K}_{Alice}$ = $\emptyset$ \\
Alice does $\pop$.($\pset$) $\rightarrow$ $\mathbb{Q}_{j}$ = ($j$, $j+1$, $\maxamount_{j,j+1}$, $\sigma_{\maxamount_{j,j+1}} ^ j$, $\sigma_{\maxamount_{j,j+1}}^{j+1}$, $tc_{j,j+1}$, $tv_{j,j+1}$,$\VKalice$)\\ 
\If{($\amount$ $\leq$ $\maxamount_{j,j+1}$)}
{

\If{($\curtime$ $<$ $tv_{j,j+1}$ )}
{

\If{($\verify_{vk_{j}}$($\maxamount_{j,j+1}$, $\sigma_{\maxamount_{j,j+1}} ^ j$) $\rightarrow$ 1 )}
{
\If{$\verify_{vk_{j+1}}$($\maxamount_{j,j+1}$, $\sigma_{\maxamount_{j,j+1}} ^ {j+1}$) $\rightarrow$ 1}
{

add $j,j+1$ to $\mathbb{K}_{Alice}$ \\
 }

\Else{
$\bcwrite$($j, j+1, \maxamount_{j,j+1}, \sigma_{\maxamount_{j,j+1} ^ {j+1}}, \VKalice$) \\
} 

}

\Else{
$\bcwrite$($j+1,j, \maxamount_{j,j+1}, \sigma_{\maxamount_{j,j+1} ^ {j}}, \VKalice$) \\
}

}

\Else{do nothing \\}

}
\Else{
$\bcwrite$($j, j+1, \amount, \maxamount_{j,j+1},  \VKalice$) \\
}

}

\For{i=1; i$\leq$ $|\mathbb{W}|_{Alice}$ ; i++}
{
\If{$\mathbb{W}_{i}$ = $\mathbb{W}_{i+1}$}

{

$\bcwrite$($\vk_{\nearrh}$, $\vk_{Alice}$, $\mathbb{W}_{i}$, $\mathbb{W}_{i+1}$) \\

}

}

}

\end{algorithm}

\begin{algorithm}[htbp]
\caption{ $\mathsf{Node  \ Leaving \ and \ Node \  Joining \ the \ DHT}$ }
\label{algo:nodejoin}
\DontPrintSemicolon
\SetAlgoLined
Let $\curl$ be the leaving node \\

\tcc{Node leaving}

$\curl$ does 
$\retrieve$($\curl$) $\rightarrow$ $\mathcal{B}_{\curl}$ \\
\If{($\mathcal{B}_{\curl}$.$\mathsf{empty}$ = False)}
{


$\curl$ does $\mathsf{FT.Delete}$($\mathcal{B}_{\curl}$) \\
}
\Else
{


\tcc{The lines 7-20 only run at the expiration of epoch $\delta$}
\For{i=1;i$\leq$ $|\mathbb{RH}|$;i++}
{



\If{($\curl$ $\in$ $\mathcal{B}_{i}$)}
{
each node $i$ does $\succlookup$($\curl$) $\rightarrow$ $\curs$ \\

\If{($\search$(${\curs}$, ${i}$)$\rightarrow$ $success$)}
         {
            do nothing \\
        }

\Else
{
}

 \If{($\open$($\vk_{\curs}$, $\vk_{i}$,\\ $lw_{\vk_{\curs}, \vk_{i}}$, $lw_{\vk_{i}, \vk_{\curs}}$) $\rightarrow$ $success$)}

 {
 channel is established \\ 
    }
    

    }
     \Else
    {

    do nothing \\ 
    }




}

}

\tcc{New node joining}
Let the joining node be $\curj$\\
$\curj$ does $H(ipaddress_{\curj})$ $\rightarrow$ ${\curj}$, $\ftcompute$(${\curj}$) $\rightarrow$ $\mathcal{B}_{\curj}$\\

$\curj$ does $\removeduplicate$($\mathcal{B}_{\curj}$) $\rightarrow$ $\mathcal{J}_{\curj}$ and $\curj$ does 

\While{($\mathcal{J}_{\curj}$.$\mathsf{empty}$ = False)}
{
$\pop$($\mathcal{J}_{\curj}$) $\rightarrow$ $i$ \\
\If{($\open$($\vk_{\curj}, \vk_{i}, lw_{\vk_{\curj},\vk_{i}}, lw_{\vk_{i},\vk_{\curj}}$) $\rightarrow$ $success$)}
{
channel is established \\

}

}


\end{algorithm}
\noindent
\underline{$\mathsf{Path \ Validation}$, Protocol \ref{algo:proof verify}}: $\mathsf{Path \ Validation}$ is the fifth protocol in $\ripd$. Alice calls this protocol once she receives the stack $\pset$ containing tuples of paths from the $\nearrh$ to all the $\rh$s in the DHT ring. In this protocol, Alice verifies the signatures of the $\rh$s contained in $\pset$ on the maximum amount that can be routed between each pair of $\rh$s that are immediate neighbors in the DHT ring. Initially, Alice pops the $\txid$ and the $\pid$ from the stack $\pset$. This stack now contains the tuples $\mathbb{Q}_{i}$, where $i$ $\in$ [$1..(|\mathbb{RH}|-1)$]. Alice adds the $\pid$ in each tuple to a list $\mathbb{W}$ that she locally maintains.
Alice retrieves each tuple and initially verifies if the amount that she intends to route is less than the maximum amount that can be routed between the $\rh$s in the tuple. If this verification fails, Alice writes the $\maxamount$, the long-term verification key of the routing helpers involved, and the $\amount$ she intends to route to the blockchain. Upon successful verification, Alice checks if the time stamp of signature validity, $tv$, is less than that of the current system time, $\curtime$. Upon successful verification, the signatures on the maximum amount created by the $\rh$s are verified. If the signature verification does not pass, Alice posts node identifiers of the malicious routing helpers involved, the $\amount$ she intends to route, the $\maxamount$ that can be transacted between the $\rh$s and the signature of the malicious $\rh$ to the blockchain. The nature of punitive actions taken against malicious parties in $\ripd$ 
may vary across PCNs. Punitive actions could include banning malicious parties temporarily or permanently, reporting them to law enforcement, etc. Describing them is beyond the scope of this paper. \\
\noindent
\underline{$\mathsf{Node \ Leaving \ and \ Node \ Joining \ the \ DHT}$, Protocol \ref{algo:nodejoin}}: This protocol handles the joining of a node in the PCN as a $\rh$ and also handles the leaving of an existing $\rh$ from the DHT ring. First, we give a description of the process of an existing $\rh$ leaving the DHT ring and follow it up with a description a node in the PCN joining as $\rh$.
 We denote the $\rh$ leaving the DHT ring by $\curl$. Initially,  
 $\curl$ calls the $\close$ function to close all the payment channels with the $\rh$s in its finger table. $\curl$ maintains local storage that stores the long-term verification keys of its neighbors. These keys are retrieved from this storage during the closing of payment channels; for brevity, we have abstracted these details in the protocol.
 Once all the payment channels have been closed, the $\rh$s in whose finger table $\curl$ was a member finds $\curl$'s successor, $\curs$. $\curs$ performs a search operation to find the $\rh$ in whose finger table $\curl$ was a member, but $\curs$ is not a member. $\curs$ then establishes payment channels with all such $\rh$s. Upon successful establishment of the payment channels, the process of an existing $\rh$ leaving the DHT ring is completed. The second part of this protocol handles the joining of a new node as a $\rh$ in the DHT ring. The node that joins the DHT ring is denoted by $\curj$. It initially finds its successor based on its node identifier in the DHT ring, denoted by $\cursucc$. Using the node identifier of $\cursucc$, $\curj$ computes the $\rh$s in its finger table and adds them to the stack $\mathcal{J}_{\curj}$. $\curj$ opens payment channels with all the $\rh$s in $\mathcal{J}_{\curj}$. This completes the process of a new node joining as a $\rh$ in the DHT.\\
 The protocol $\mathsf{Routing \ Payment}$, Protocol \ref{algo:routepay} handles the routing of payment between Alice and Bob using HTLCs \cite{htlc}. The steps for this protocol are self explanatory and we give the protocol and its full description in Appendix \ref{sec:appendixproto}.

%% file: impl_full.tex
\section{Experimental Evaluation}
\label{sec:impl}
In this section, we explain our dataset collection, experimental setup and the results of our evaluation.
\subsection{Dataset and Simulation Setup}
In $\ripd$, we use the transaction data from Ripple for our experimental evaluation. Transaction data about the most popular PCN, the Lightning Network, in particular, the data about the number and the amount of transactions is not publicly available. Due to this, 
and the fact that Lightning and Ripple are the only PCNs in use currently, we use transaction data from Ripple for our experiments. $\ripd$, however, can be deployed on Lightning Network without any modification to the underlying structure of Lightning Network. The only overhead that $\ripd$ causes when deployed on Lightning Network and Ripple is the opening of payment channels in Lightning Network (called trustlines in Ripple~\cite{trustlines}) by the $\rh$s with the entries in their finger tables and the creation and verification of pseudonymous identities for every node in the PCN. 
\begin{table}[H]

      \caption{Number of cryptographic operations performed/TX. Legend: $\mathbb{LM}$: landmark, $d$: size of the hash digest used in the DHT, $T$: number of cryptographic operations performed outside the DHT.}
  \label{tbl:crytooperations}
      \centering

          \begin{tabular}
{|c|c|c|c|}
\hline
\multicolumn{1}{|c|}{\multirow{2}{*}{Operations}} & \multicolumn{3}{c|}{Protocols}                                        \\ \cline{2-4} 
\multicolumn{1}{|c|}{}                            & \multicolumn{1}{l|}{$\ripd$}    & \multicolumn{1}{l|}{Blanc} & SW       \\ \hline
Signing                                           & \multicolumn{1}{l|}{$1+T$}        & \multicolumn{1}{l|}{13}    & $8|\mathbb{LM}|+1$ \\ \hline
Verification                                      & \multicolumn{1}{l|}{$|\mathbb{RH}|+T$}     & \multicolumn{1}{l|}{12}    & $7+|\mathbb{LM}|$   \\ \hline
Hash                                           & \multicolumn{1}{l|}{$3+T$}        & \multicolumn{1}{l|}{7}     & 0        \\ \hline
Encryption                                        & \multicolumn{1}{l|}{$T$}        & \multicolumn{1}{l|}{7}     & 0        \\ \hline
Decryption                                        & \multicolumn{1}{l|}{$T$}        & \multicolumn{1}{l|}{6}     & 0        \\ \hline
$\lookup$                             & \multicolumn{1}{l|}{$O(\log d)$} & \multicolumn{1}{l|}{0}     & 0        \\ \hline

$\ftcompute$                            & \multicolumn{1}{l|}{$\log d$} & \multicolumn{1}{l|}{0}     & 0        \\ \hline
\end{tabular}
        
    
\end{table}
For the experiments, we collected transaction data from the Ripple network from 01-01-2021 to 12-31-2021~\cite{ripple}. We chose to collect Ripple data due to the fact that Ripple's XRP token, has the sixth largest capitalization for a cryptocurrency and Ripple's market cap is the largest among all payment channel networks~\cite{ripplemarketcap}. 
We used the Ripple API \cite{rippleapi} to collect all the ``Payment'' transactions that were recorded on the Ripple ledger during the aforementioned time period. We only consider the ``Payment'' transactions since they are path-based transactions that involve several intermediate nodes between the $\send$ and $\rec$. These transactions were recorded in several different currencies. Direct transactions between a sender and receiver pair which do not involve intermediaries were excluded from our collected data.  We collected a total of 15,634,656 path-based transactions. We pre-processed the collected data to remove two types of anomalies that we have noticed: invalid currencies, and incomplete hash digests of transactions. Once the transaction data was collected and pruned, we created a directed graph using the Ripple APIs~\cite{trustlines} for every month from January 2021 to December 2021. We removed the edges with negative and zero link weights and converted all the link weights to USD. This gave us a graph of 225,264 nodes and 1,717,347 edges which was used in our experiments. All our experiments were run on a single machine equipped with AMD $\textsuperscript{TM}$ EPYC processor (64 bit architecture) with 16 cores and 512 GB of RAM and a clock speed of 3.2 GHz. The code for all the routing algorithms was written in Python 3.8 and the NetworkX library~\cite{networkx} was used for simulations.
\subsection{Evaluation And Results}
We implement and experimentally compare $\ripd$ with several other comparable routing algorithms, specifically with SilentWhispers \cite{Malavolta2016SilentWhispersES}, SpeedyMurmurs \cite{roos2017settling} and Blanc \cite{blanc} and show our results in Table \ref{tbl:allresults}. For all the experimental settings, we set the $\rh$s for $\ripd$ and Blanc~\cite{blanc}, and the number of landmarks for SilentWhispers~\cite{Malavolta2016SilentWhispersES} and SpeedyMurmurs~\cite{roos2017settling} to eight. These routing helpers/landmarks were picked as the nodes with the highest out-degree in the graph. 
In our experiments, we constructed the graph for each month in 2021 and routed the transactions accordingly. By doing so, we capture the growth of the Ripple network over the year through our graph. Our experiments thus simulate the network's evolving, dynamic nature. 
\begin{table*}[h!]
\caption{Performance of different pathfinding and routing protocols. Legend: PL: path length, 1-SCC: one large strongly connected component in the PCN graph, $k$-SCC: $k$ strongly connected components. Metrics: success ratio (higher is better), mean path length (lower is better), pathfinding time (lower is better), routing time (lower is better).}
\centering
\label{tbl:allresults}
\begin{tabular}
{|p{2.50cm}|p{2.50cm}|p{2.25cm}|p{2.25cm}|p{2.20cm}|}
\hline
Protocols                         & Success ratio  & Mean PL (hop-count) & Pathfinding time (sec) & Routing time (millisec)\\ \hline\hline
$\ripd$ 1-SCC (single graph) & 98.85 $\pm$ 0.027                       & 6.64 $\pm$ 0.287                                 & 31.24169 $\pm$ 56.144           & 3.126 $\pm$ 0.0254    \\ \hline
$\ripd$ $k$-SCC (several disjoint graphs)   & 98.73 $\pm$ 0.148                                & 7.31 $\pm$ 0.295                                  & 31.24289  $\pm$ 56.147         & 3.165 $\pm$ 0.03    \\ \hline
SM 1-SCC (single graph)~\cite{roos2017settling} & 98.23 $\pm$ 1.64  & 4.21 $\pm$ 0.245  & 12460.0688 $\pm$ 11089.897 & 3500 $\pm$ 0.042  \\ \hline
SW 1-SCC (single graph)~\cite{Malavolta2016SilentWhispersES}  & 94.63 $\pm$ 7.08  & 7.87 $\pm$ 0.387  & 51707.89  $\pm$ 11651.260 & 225000  $\pm$ 630 \\
\hline
Blanc 1-SCC (single graph)~\cite{blanc} & 97.92 $\pm$ 0.029 & 10.983 $\pm$ 0.754  & 56344.229 $\pm$ 186149.669  &  391164 $\pm$ 725.458 \\
\hline
\end{tabular}
\end{table*}
The graphs for each month constructed from the Ripple data were disjoint. Hence we extracted the largest strongly connected component for each month's graph and routed the transactions (involving the USD currency) by selecting the sender-receiver pairs and routing helpers from that component. This is the ``$\ripd$-1-SCC'' setting in the Table \ref{tbl:allresults}. In order to demonstrate the effectiveness of having the routing helpers connected via a DHT such as Chord (which is the central idea of $\ripd$), we extracted the top eight (by node count) strongly connected components for each month.
We selected one routing helper from each strongly connected component (based on the highest out-degree), connected them via a Chord ring, and all the transactions that were recorded for the USD currency in our dataset were routed through these $\rh$s. We randomly sampled the $\send$ and $\rec$ from the dataset to ensure that no $\send$ and $\rec$ pair is from the same SCC. This is the ``$\ripd$-k-SCC'' setting in Table~\ref{tbl:allresults}. 

We measured a total of four metrics for each month: 1) the transaction success ratio, which is the ratio of the number of transactions successful to the total number of transactions routed, 2) the average path length found, which is the total number of hops between the sender and receiver, 3) pathfinding time, which is the time taken to find a path between the sender and the receiver, and 4) routing time, which is the time taken to route the payment (nodes adjusting link weights) along the path from the sender to the receiver. We computed the mean of each metric across all twelve months  along with the standard deviation. A total of 52,943 transactions which is the number of transactions that took place with the currency as USD during 2021-2022 on the Ripple ledger 
were routed concurrently. 

Message passing between nodes in the course of a routing protocol in a p2p network such as PCN is an implementation-specific scenario that depends on the network in which the routing protocols are deployed. For instance, the Lightning network uses the in-built IPv4 or IPv6 connection that exists between the nodes in the PCN for message passing. For more information on this, we refer the reader to \cite{message}. Ripple uses the Ripple Protocol Consensus Algorithm (RPCA) \cite{rpca}, which internally handles message passing. Similar to these, in our simulations for $\ripd$, we use the NetworkX library, which handles message passing internally.

We used Dijkstra's shortest path algorithm to simulate the pathfinding outside the DHT in $\ripd$. In $\ripd$, the number of edges $|E|$ is significantly lower than $|V|^{2}/\log |V|$ for the PCN graph $G(V,E)$, hence we implemented the priority queue for Dijkstra's algorithm using a binary heap~\cite{dasgupta2008algorithms}. For $\ripd$, the shortest path (in terms of the hop count) was chosen to route the payment from $\send$ to $\nearrh$, $\nearrh$ to other $\rh$s in the DHT ring, and $\erh$ to $\rec$. If multiple paths with the same hop count were present, the paths with the highest liquidity were chosen. If the hop-count and the liquidity between paths were the same, then a path was randomly chosen.

For the simulation of SilentWhispers \cite{Malavolta2016SilentWhispersES}, the number of landmarks was chosen as eight, and the landmarks were chosen as the nodes with the highest out-degree. Unlike $\ripd$, SilentWhispers cannot be applied to disjoint graphs, since it uses BFS (breadth first search) to find a path between the $\send$ and $\rec$. Even though BFS is asymptotically more efficient than Dijkstra's algorithm, the routing time and pathfinding times are significantly higher than $\ripd$, since the overhead contributed by the number of cryptographic operations (signing and verification) is very high. Besides, unlike $\ripd$, ~\cite{Malavolta2016SilentWhispersES} offers no support for concurrent transactions. In $\ripd$ the path length inside the DHT ring is always 3 hops since we have chosen a total of 8 routing helpers. 

SpeedyMurmurs~\cite{roos2017settling} uses an embedding-based routing mechanism called VOUTE \cite{roos2016voute}. VOUTE uses a BFS-based approach to compute the embedding of all the nodes in the network with respect to their distances from the landmark. Apart from this, the BFS needs to be run by all the landmarks when a new node joins or an existing node leaves the network, which gives this protocol a high stabilization overhead. Hence the pathfinding time for this is higher than $\ripd$. However, since there are no cryptographic operations involved, the pathfinding time is lower than that of SilentWhispers.
The routing time for this protocol is also less than that of SilentWhispers and Blanc since the actual payment routing does not involve any cryptographic operations.  This protocol also offers no privacy guarantees unlike $\ripd$, where we offer $\send$, $\rec$, and transaction privacy. 

The routing protocol Blanc~\cite{blanc}, was simulated with  the number of $\rh$s chosen as eight similar to all the other protocols. It needs two $\rh$s between the $\send$ and the $\rec$, one picked by the sender and the other picked by the $\rec$. The routing/pathfinding time is very high in this protocol in comparison to the other protocols, since the pathfinding phase uses broadcasting in three segments: $\send$ to $\rh$1, $\rh$1 to $\rh$2, $\rh$2 to $\rec$. Blanc does not address the issue of multiple SCCs. The routing time is higher than $\ripd$, SilentWhispers, and SpeedyMurmurs, because Blanc involves the creation of pair-wise contracts (after the pathfinding phase) between nodes involved along the path attesting to the amount that will be transacted. Table \ref{tbl:crytooperations} represents the number of cryptographic operations performed by each routing protocol per every transaction being routed. The transaction processing time for fiat currencies varies from a couple of hours to a couple of days depending on the geographical location of the sender and the receiver \cite{zelle,xoom,xoomlimit}. Our experimental evaluation shows that the average pathfinding time for $\ripd$ is 31 seconds and the average routing time is 3 milliseconds across 52,000 transactions that were recorded for an year. This shows the efficiency of $\ripd$ in particular. 

\textbf{Setup time analysis}: The setup time for Blanc and SilentWhispers is equal (4.565 seconds) since their setup involves only the creation of the signing and verification keys for the nodes and the creation and verification of pseudonymous identities. This step can be parallelized. The setup time for SpeedyMurmurs is the highest, $\approx$ 5.9 hours since it involves computation of the embedding coordinates of nodes in the PCN. The one-time setup time for $\ripd$ is also significantly high, $\approx$ 4.12 hours since it involves an additional setup for the establishment of the DHT ring and the creation and verification of a node's long-term and pseudonymous identities.

\textbf{Tradeoffs}:
$\ripd$ introduces a delay whenever the finger table of an $\rh$ needs to be updated in the event of another $\rh$ joining or leaving the DHT ring. In a DHT, at most $log(m)$ entries in a node's finger table can be distinct where $m$ is the size of the digest obtained by the hashing the node's IP address. 
Updating an existing $\rh$'s (when another $\rh$ leaves the DHT) finger  table or a newly joined $\rh$ creating a finger table involves the existing $\rh$ or the newly joined $\rh$ opening payment channels with their finger table entries. In the worst case, a $\rh$ night need to open $log(m)$ payment channels. This introduces a delay of $\alpha$, where $\alpha$ is the time to process payment channel openings depending on the blockchain on which $\ripd$ will be deployed. The payment channels between $\rh$s and their finger table entries can be opened in parallel. For BTC this delay varies from 60 to 90 minutes.

The payment channels opened by the $\rh$s can be used for routing multiple transactions, which amortizes the delay $\alpha$ over several thousands of transactions. 
Our evaluations show this delay being amortized over 52,000 transactions with an average pathfinding time of thirty one seconds and an average routing time of three milliseconds. In other words, the setup time of $\ripd$, which is close to four hours is amortized over routing 52,000 transactions being routed in three milliseconds. 
In addition to this, updating the maximum amount at the end of each time epoch ($\delta$) introduces a delay of $\beta$, which is the time taken for a $\rh$ and a corresponding finger table entry to sign the maximum amount that can be transacted between them. This makes the total delay introduced by $\ripd$ as ($\alpha+\beta$).

%% file: secanalysis_full.tex
\section{$\ripd$ Security Analysis}
\label{sec:security}
 
 In this section, we provide a formal analysis of $\ripd$. We define an ideal functionality $\fraced$, that consists of six functionalities: $\finit$, $\fdht$, $\faux$ $\ffindpath$, $\fpayment$, and $\fhtlc$, and 2 helper functionalities $\fsigverify$ \cite{canetti2004universally}, and $\fpcn$ \cite{malavolta2017concurrency}. 
\begin{theorem}

\label{thm:uctheorem}
Let $\fraced$ be an ideal functionality for $\ripd$. Let $\adversary$ be a probabilistic polynomial-time (PPT) adversary for $\ripd$, and let $\simulator$ be an ideal-world PPT simulator for $\fraced$. $\ripd$ UC-realizes $\fraced$ for any PPT distinguishing environment $\mathcal{Z}$.

\end{theorem}



\begin{figure}[H]
\caption{$\finit$ Ideal Functionality}
\label{fig:finit}
\fbox{\begin{minipage}{37em}
\flushleft
\begin{itemize} 
\item \textbf{Key Generation}: 
 Upon receiving the tuple ($\keygenLT$, $\idsender_{i}$)  from  node $i$ or $\simulator$, $\finit$ forwards it to $\fsigverify$. Upon receiving ($\longverificationkey$, $\idsender_{i}$, $v_{i}$) and ($\tempverificationkey$, $\idsender_{i}$, $V_{i}$), $\finit$  stores the  tuple ($\idsender_{i}$, $v_{i}$, $V_{i}$) 
in the table $\utable$ and sends the tuple ($\longverificationkey$, $\idsender_{i}$, $v_{i}$) and ($\tempverificationkey$, $\idsender_{i}$, $V_{i}$) to the node and the simulator $\simulator$. Upon receiving the tuple ($\sign$, $\idsender_{i}$, $V_{i}$) from node $i$ or $\simulator$, $\finit$ sends the tuple ($\sign$, $\idsender_{i}$, $V_{i}$) to $\fsigverify$. If $\fsigverify$ responds with  ($\signature$, $\idsender_{i}$, $V_{i}$, $\sigma_{V_{i}}$), $\finit$ updates the corresponding entry in $\utable$ to  ($\idsender_{i}$, $\cdot$, $\cdot$, $\sigma_{V_{i}}$ ) and sends the tuple ($\signature$, $\idsender_{i}$, $V_{i}$, $\sigma_{V_{i}}$) to the node and also to  $\simulator$.  Else $\finit$ returns $\bot$.
\end{itemize}
\begin{itemize}
\item \textbf{Identity Verification}: Upon receiving the tuple ($\neighbors$, $\mathbb{I}_{i}$) from node $i$ or $\simulator$, $\finit$ sends the tuple ($\tempverificationkey$, $V_{i}$) to all the nodes in $\mathbb{I}_{i}$. Upon receiving the tuple ($\verification$, $\idsender_{i}$, $V_{i}$, $\sigma_{V_{i}}$, $v_{i}$) from nodes in $\mathbb{I}_{i}$ or $\simulator$, $\finit$ sends the tuple ($\verification$, $\idsender_{i}$, $V_{i}$, $\sigma_{V_{i}}$, $v_{\idsender}$) to $\fsigverify$. Upon receiving ($\verification$, $\idsender_{i}$, $V_{i}$, $f$) from $\fsigverify$, $\finit$ updates the $\utable$ with  ($\idsender_{i}$, $\cdot$, $\cdot$ ,$\cdot$, $f$) and sends the tuple ($\verification$, $\idsender_{i}$, $V_{i}$, $f$) to the nodes in $\mathbb{I}_{i}$ and this tuple is also sent to the $\simulator$. 
\end{itemize}
\end{minipage}}
\end{figure}
We provide a formal analysis of $\ripd$ in the Universal Composability (UC) framework \cite{DBLP:conf/focs/Canetti01}. 
The notion of UC security is captured by the pair of definitions below:

\begin{definition}
(UC-emulation \cite{DBLP:conf/focs/Canetti01}) Let $\pi$ and $\phi$ be probabilistic polynomial-time (PPT) protocols. We say that $\pi$ UC-emulates $\phi$ if for any PPT adversary $\adversary$ there exists a PPT adversary $\simulator$ such that for any balanced PPT environment $\mathcal{Z}$ we have 
\begin{center}
    $\mathsf{EXEC}_{\phi, \simulator, \mathcal{Z}}$ $\approx$ $\mathsf{EXEC}_{\phi, \adversary, \mathcal{Z}}$
\end{center}
\end{definition}
\begin{definition}
    (UC-realization \cite{DBLP:conf/focs/Canetti01}) Let $\mathcal{F}$ be an ideal functionality and let $\pi$ be a protocol. We say that $\pi$ UC-realizes $\mathcal{F}$ if $\pi$ UC emulates the ideal protocol for $\mathcal{F}$.
\end{definition}
We define an ideal functionality $\fraced$, that consists of six functionalities: $\finit$, $\fdht$, $\faux$, $\ffindpath$, $\fpayment$, and $\fhtlc$. We also use two helper functionalities, $\fsigverify$ \cite{canetti2004universally} and $\fpcn$ \cite{malavolta2017concurrency}. We now describe these ideal functionalities and  give the proof of theorem \ref{thm:uctheorem}. \\
1) \underline{{ $\finit$ functionality}}: This functionality depicted in Figure~\ref{fig:finit}. This functionality handles the generation and verification of identities for all the nodes in the PCN. This functionality performs two operations, \textbf{Key Generation} and \textbf{Identity Verification}. In the Key Generation, the $\finit$ functionality generates the long-term and temporary signing and verification key pairs using the $\fsigverify$ functionality. The purpose of creating a temporary identity is to hide the real identity of a node from its non-neighboring nodes in the network.
Once the identities are generated, $\finit$ also handles the verification of the node's temporary identity by its neighbors, which is the identity verification phase. \\
2) \underline{{$\fdht$ functionality}}: This functionality is shown in Figure~\ref{fig:fdht} and handles the joining of an existing node in the PCN as a routing helper in the DHT ring. For any node that joins the DHT ring, this functionality performs three operations. 
\begin{figure}[H]
\caption{$\fdht$ Ideal Functionality}
\label{fig:fdht}
\fbox{\begin{minipage}{37em}
\flushleft
\begin{itemize}
\item \textbf{Payment channel opening}: Upon receiving the tuple ($\mathsf{Open}$, $c_{\langle \curk, \curi \rangle}, v, \curi, t, f$) from the $\rh$ $\idsender_{\curk}$ for each node $\idsender_{\curi}$ $\in$ $\jset_{\curk}$ or $\simulator$, $\fdht$ sends the tuple to $\fpcn$ functionality.
 Upon receiving the tuple ($c_{\langle \curk, \curi \rangle}$,$v, t, f$) from $\fpcn$ for each $\rh$ $\idsender_{\curi}$ $\in$ $\jset_{\curk}$, $\fdht$ sends this tuple to $\idsender_{\curk}$ and the simulator $\simulator$. Upon receiving the tuple ($\mathsf{Authorize}$, $c_{\langle \curk, \curi \rangle}$,$v, t, f$) from $\idsender_{\curk}$ for each $\rh$ $\idsender_{\curi}$ $\in$ $\jset_{\curk}$, $\fdht$ sends this tuple to $\fpcn$. Upon receiving $h_{i}$ from $\fpcn$ for each $\rh$ $\idsender_{\curi}$ $\in$ $\jset_{\curk}$ or $\simulator$, $\fdht$ sends $h_{i}$ for each $\curi$ in $\jset_{\curk}$ to both $\idsender_{\curi}$ and $\idsender_{\curk}$ and also the simulator $\simulator$ .

\end{itemize}
\flushleft
\begin{itemize}
\item \textbf{Maximum amount computation}: Upon receiving the tuple ($\mathsf{Maximum~Amount}$, $\maxamount_{k,i}$, $\sigma_{k,i}^k$) from $\idsender_{\curk}$ for all nodes $i$ $\in$ $\mathcal{J}_{k}$ or $\simulator$, $\fdht$ checks if it received a tuple ($\mathsf{Maximum~Amount}$, $\maxamount_{k,i}$, $\sigma_{k,i}^i$) from  $i$ or $\simulator$.
If yes, $\fdht$ adds $(\maxamount_{k,i}, \sigma_{k,i}^k, \sigma_{k,i}^i, tc_{k,i}, tv_{k,i})$, the values of $tc_{i,k}$ and $tv_{i,k}$ $\leftarrow$ $\mathbb{T}$ to list $\mathcal{M}_k$, and updates tuple $(sid_{k}, \cdot, \mathcal{M}_k)$, else it returns $\bot$.  Once the time epoch $\delta$ has expired, $\fdht$ checks if its has received the tuple ($\mathsf{Maximum~Amount} ^ \prime$, $\maxamount_{k,i} ^ \prime$, $\sigma_{\maxamount_{k,i} ^ \prime}^{k}$) from  $k$ for all nodes $i$ $\in$ $\mathcal{J}_{k}$ or $\simulator$. If yes, $\fdht$ checks if it has received a tuple ($\mathsf{Maximum~Amount} ^ \prime$, $\maxamount_{k,i} ^ \prime$, $\sigma_{\maxamount_{k,i} ^ \prime}^{i}$) for each node $\curk$ or $\simulator$. If yes, $\fdht$ updates the list $\mset_{k}$ with $\mathsf{Maximum~Amount} ^ \prime$, $\maxamount_{k,i} ^ \prime$, $\sigma_{\maxamount_{k,i} ^ \prime}^{i}$, $tc_{i,k}^\prime$, $tv_{i,k}^\prime$). The values of $tc_{i,k}^\prime$ and $tv_{i,k}^\prime$ $\leftarrow$ $\mathbb{T}$.
\end{itemize}

\flushleft
\begin{itemize}
\item \textbf{Node leaving}: Upon receiving the tuple ($\mathsf{DHT~Leave}$, $\idsender_{\curk}$) from a $\rh$ or $\simulator$ $\idsender_{\curk}$, the $\fdht$ functionality deletes the entry ($\idsender_{\curk}$, $\jset_{\curk}$, $\mset_{\curk}$) from the $\rtable$.
\end{itemize}
\end{minipage}}   
\end{figure}
\textbf{Finger table computation}: $\fdht$ chooses a time epoch $\delta$ $\leftarrow$ $\mathbb{R} ^ {+}$ and picks a hash function whose digest size is $m$ $\leftarrow$ $\mathbb{Z}^{+}$. Upon receiving the tuple ($\mathsf{DHT~Join}$, $\idsender_{k}$) from a  node $k$ in the PCN or the simulator $\simulator$, $\fdht$  computes  $l_{j}$ = (${k}$ + $2 ^ {(j-1)}$ $\mod$ $m$), $\forall j \in [1, \cdots, m]$. $\fdht$ retrieves the $\rh$ whose node identifier is the smallest and is less than or equal to the value of $l_{j}$ and this node identifier is denoted by $\mathsf{Succ}$($l_{j}$). Each value of $l_{j}$ is added to the list $\mathcal{J}_{k}$. $\fdht$ then stores the tuple ($\idsender_{k}$, $\mathcal{J}_{k}$) in the $\rtable$ and sends the tuple ($\fingertable$, $\mathcal{J}_{k}$) to $k$. This tuple is also sent to the simulator $\simulator$. 
The \textbf{Finger table computation} populates the finger table of the newly joined node as the $\rh$. Upon successful completion of the finger table generation, this functionality stores it locally and sends this finger table to the newly joined $\rh$.
In the case of a simulator, $\simulator$ simulating this functionality, the finger table is sent to the simulator. This phase also chooses the time epoch, $\delta$. Upon the expiry of this time epoch, each $\rh$ in the DHT ring computes the maximum amount, $\maxamount$ that it can transact to the nodes in its finger table and also generates the appropriate signatures. The second phase is the \textbf{Payment channel opening}. Upon the successful creation of a finger table for a newly joined $\rh$, this functionality opens payment channels between the newly joined $\rh$ and its finger table entries. The $\fdht$ functionality makes use of the $\fpcn$ functionality for the payment channel opening. Upon successful opening of the payment channel, the identifier $h$ for each channel opened is sent to the newly joined $\rh$.  In the case of a simulator, $\simulator$ simulating this functionality, the identifier $h$ is sent to the simulator. The third operation carried out by this functionality is the \textbf{Maximum amount computation}. Once the payment channels have been opened for each node in the newly joined $\rh$'s finger table, this functionality computes the maximum amount that the $\rh$ $i$ can transact with each node in its finger table, $k$ denoted by $\maxamount_{i,k}$. This computation is done upon the expiry of the time epoch value $ \delta$. Apart from these, this functionality also handles the leaving of a $\rh$ from the DHT ring. Once a node decides to leave the DHT ring, the $\fdht$ functionality simply deletes the corresponding $\rh$ from its local storage.\\
3)\underline{{$\ffindpath$ functionality}}: This functionality shown in Figure~\ref{fig:ffindpath} handles the pathfinding between the $\nearrh$, and all the other $\rh$s in the DHT ring that can route the amount, $\amount$ specified by the $\send$.
This functionality performs 2 operations. \textbf{Find path phase-1} and \textbf{Find path phase-2} In phase-1, this functionality finds paths from  $\nearrh$ to all the nodes in its finger table that can route the $\amount$ specified by the $\send$. Since $\nearrh$ and the nodes in this phase are connected by direct payment channels, the functionality simply checks if the $\amount$ $<$ $\maxamount_{\nearrh,k}$ that can be routed by $\nearrh$ to each of its finger table entries (where $k \in \bset_{\nearrh}$) and adds the $\rh$s that can route the $\amount$ specified by the $\send$ to a stack $\mathbb{P}$.
In phase-2, the functionality finds paths between $\nearrh$ and all the $\rh$s in the DHT ring that are not in the finger table of $\nearrh$ and can route the $\amount$ specified by the $\send$. The $\nearrh$ finds path to each such $\rh$ by choosing the node that is closest to the destination from its finger table. Upon completion of these 2 phases, the $\ffindpath$ functionality sends the stack of paths, $\pset$ to the $\nearrh$. \\
4)\underline{{$\fpayment$ functionality}}:  This functionality is depicted in Figure \ref{fig:fpayment} and handles the payment mechanism between the $\send$ and the $\rec$.
In $\ripd$ the payment is completed using HTLCs. This functionality performs two operations. \textbf{Routing helper and final path selection} and \textbf{Payment routing}. The routing helper and final path selection handles the selection of an $\erh$ by the $\rec$. The functionality sends the tuple containing all the $\erh$ in all the paths received by the $\send$ from the $\ffindpath$ functionality to the $\rec$. The $\rec$ picks the $\erh$ that is closest to him in terms of hop count and the $\fpayment$ functionality selects the path that contains the $\rh$ picked by the $\rec$ as the $\erh$. This completes the routing helper and final path selection phase. The payment routing phase then completes the payment procedure using HTLCs. The preceding node communicates the preimage to the $\fpayment$ functionality and the succeeding node communicates the digest. These values are forwarded to the $\fhtlc$ functionality to complete the payment.
\begin{figure}[H]
\caption{$\fpayment$ ideal functionality} 
\label{fig:fpayment}
\fbox{\begin{minipage}{37em}
\begin{itemize}
\item \textbf{Routing helper and final path selection:} Upon receiving the tuple ($\mathrm{list \ of \erh}$, $\tset$, $\txid$, $\VK_{Alice}$) from Alice, $\fpayment$ retrieves the $\txid$ and the stack of paths $\pset$ from the $\txtable$ =  ($\idsender_{Alice}$, $\txid$, $\cdot$, $\pset$, $\cdot$, $\cdot$, $\cdot$, $\cdot$). $\forall$ $i$ $\in$ [$1 \cdots 4\times (log|\mathbb{RH}|)$], where $4 \times log(|\mathbb{RH}|)$ is the size of the stack $\pset$, $\fpayment$ retrieves the entry indexed by $i$, such that, $i \mod 4 =0$ and adds the entry, which is the $\erh$ to a list $\mathbb{M}$ that it locally maintains. $\forall$ $k$ $\in$ $\mathbb{M}$ $\fpayment$ checks if $k$ $\in$ $\tset$. If all these checks pass, $\fpayment$ updates the $\txtable$ to store ($\idsender_{Alice}$, $\txid$, $\amount$, $\pset$, $\tset$) and sends the tuple ($\mathrm{list \ of \erh}$, $\tset$, $\txid$, $\VK_{Alice}$) to Bob or $\simulator$. Else it returns a $\bot$. Upon receiving the tuple ($\mathrm{\erh, \erh_{Bob}, \txid, \VK_{Bob}}$, $Y$, $X$) from Bob or $\simulator$, $\fpayment$ checks if $\erh_{Bob}$ $\in$ $\tset$. $\fpayment$ also checks if $\txid$ $\in$ $\txtable$ = ($\cdot$, $\txid$, $\cdot$, $\cdot$, $\cdot$, $\cdot$, $\cdot$) and $\VK_{Bob}$ $\in$ $\utable$ = ($\idsender_{Bob}$, $\cdot$, $\VK_{Bob}$, $\cdot$, $\cdot$).
If all these checks pass, $\fpayment$ updates to the $\txtable$ to store ($\idsender_{Alice}$, $\txid$, $\amount$, $\pset$, $\tset$, $\erh_{Bob}$, $Y$, $X$) and sends the tuple  ($\mathrm{\erh, \erh_{Bob}, \txid, \VK_{Bob}}$, $Y$) to Alice or $\simulator$. Else it returns a $\bot$. Upon receiving the tuple ($\mathrm{Final \ Path}$, $\VK_{i}$, $\VK_{i+1}$) from each node $i$ along the path of $\txid$ from Alice to Bob or $\simulator$, the $\fpayment$ checks if $\VK_{i}$ $\in$ $\utable$ = ($\idsender_{i}$, $\cdot$, $\VK_{i}$, $\cdot$, $\cdot$), checks if  $\VK_{i+1}$ $\in$ $\utable$ = ($\idsender_{i+1}$, $\cdot$, $\VK_{i+1}$, $\cdot$, $\cdot$). If these checks pass, $\fpayment$ then retrieves the $\pset$, $\mathbb{O}$, $\mathbb{O}^\prime$ from the $\txtable$ = ($\idsender_{Alice}$, $\cdot$, $\mathbb{O}$, $\mathbb{O}^\prime$, $\cdot$, $\pset$, $\cdot$, $\cdot$, $\cdot$, $\cdot$). For every pair of consecutive nodes in  $\pset$, $\mathbb{O}$ and $\mathbb{O}^\prime$, $\fpayment$ checks if $i,i+1$ = $\mathbb{P}[i],\mathbb{P}[i+1]$ or $i,i+1$ = $\mathbb{O}[i],\mathbb{O}[i+1]$ or $i,i+1$ = $\mathbb{O}^\prime[i],\mathbb{O}^\prime[i+1]$ If yes, $\fpayment$ sends the tuple ($\mathrm{Final \ Path}$, $\txid$, $\VK_{i}$, $\VK_{i+1}$) to node $i+1$ or $\simulator$.

\item \textbf{Payment routing:} For every pair of consecutive nodes $i,i+1$ in the path along the $\txid$ from Bob to Alice or from $\simulator$, when $\fpayment$ receives the tuple ($\mathsf{HTLC \ tuple}$, $ \vk_{i}$, $\vk_{i+1}$, $\txid$, $\amount$, $Y$, $X$) from $i+1$, $\fpayment$ retrieves the $\txid$, $\mathbb{O}$, $\mathbb{O}^\prime$, $\pset$ from the $\txtable$ = ($\idsender_{Alice}$, $\txid$, $\mathbb{O}$, $\mathbb{O}^\prime$, $\cdot$, $\pset$, $\cdot$, $\cdot$, $\cdot$, $\cdot$) respectively and checks if the $\txid$ sent by the pair of consecutive nodes $\in$ $\txtable$. For every pair of consecutive nodes in  $\pset$, $\mathbb{O}$ and $\mathbb{O}^\prime$, $\fpayment$ checks if $i,i+1$ = $\mathbb{P}[i],\mathbb{P}[i+1]$ or $i,i+1$ = $\mathbb{O}[i],\mathbb{O}[i+1]$ or $i,i+1$ = $\mathbb{O}^\prime[i],\mathbb{O}^\prime[i+1]$. If all these checks pass, this tuple is forwarded to $\fhtlc$ by the $\fpayment$. Else a $\bot$ is returned.  Upon receiving a message ($\mathsf{Success}$) or $\bot$ from the $\fhtlc$, these messages are forwarded to the pair of consecutive nodes by $\fpayment$ or $\simulator$.
\end{itemize}
\end{minipage}}
\end{figure} 
\begin{figure}[htbp]
\caption{$\ffindpath$ Ideal Functionality}
\label{fig:ffindpath}
\fbox{\begin{minipage}{37em}
\flushleft
\begin{itemize}
\item \textbf{Find path phase-1}: Upon receiving the tuple ($\txid$, $\mathsf{Verified}$, $\idsender_{Alice}$, $\amount$, $f$, $\idsender_{\nearrh}$) from Alice or $\simulator$ , $\ffindpath$ stores the ($\idsender_{Alice}$, $\txid$, $\amount$, $\cdot$, $\cdot$, $\cdot$, $\cdot$, $\cdot$, $\cdot$, $\cdot$) in the $\txtable$ and checks the value of $f$. If $f$ = 0 or If $f=\phi$, $\ffindpath$ returns a $\bot$ and aborts. Else if $f=1$, $\ffindpath$ retrieves $\jset_{\nearrh}$ from the tuple ($\idsender_{\nearrh}$, $\jset_{\nearrh}$, $\mset_{\nearrh}$) from $\rtable$ and sends the tuple ($\txid$, $\amount$, $\sigma_{\amount}$, $\VKalice$) to all $\idsender_i$ $\in$ $\jset_{\nearrh}$. Upon receiving the tuple ($\mathsf{Verified}$, $\idsender_{Alice}$, $\amount$, $f$) from $\fsigverify$, $\ffindpath$ checks the value of $f$. If $f=0$ or $f=\phi$, $\ffindpath$ returns a $\bot$ and aborts. If $f=1$,  $\ffindpath$ retrieves the $\idsender_{j}$ of $\rh$s $\notin$ $\jset_{\nearrh}$ but present in $\rtable$ and adds them to them to an empty stack $\mathbb{W}$. For each $\rh$ $\curi$ $\in$ $\jset_{\nearrh}$ ,
$\ffindpath$ retrieves the values of $\maxamount_{\nearrh,i}$, $\sigma_{\maxamount_{\nearrh,\curi}}^{\nearrh}$, $\sigma_{\maxamount_{\nearrh,\curi}}^{\curi}$ from the list $\mset_{\nearrh}$ ($\maxamount_{\nearrh,i}$, $\sigma_{\maxamount_{\nearrh,\curi}}^{\nearrh}$ , $\sigma_{\maxamount_{\nearrh,\curi}}^{\curi}$, $\cdot $, $\cdot$) which is in the $\rtable$ ($\nearrh_{\nearrh}$, $\cdot$, $\mset_{\nearrh}$), such that $\maxamount_{\nearrh,\curi}$ $\geq$ $\amount$, and samples a $\pid_{\curi}$ $\sample$ $\{0,1\}^\lambda$ and adds ($\txid$, $\pid_{\curi}$, $\nearrh$, $\curi$, $\maxamount_{\nearrh,\curi}$, $\sigma_{\maxamount_{\nearrh,\curi}}^{\nearrh}$, $\sigma_{\maxamount_{\nearrh,\curi}}^{i}$, $tc_{\nearrh, \curi}$, $tv_{\nearrh, \curi}$) to an empty stack $\mathbb{P}$.
\end{itemize}
\flushleft
\begin{itemize}
\item \textbf{Find path phase-2}: $\ffindpath$ pops each element $\idsender_{\curp}$, where $\idsender_{\curp}$ from the stack $\mathbb{W}$.  $\ffindpath$ retrieves $\jset_{\nearrh}$ from $\rtable$ ($\idsender_{\nearrh}$, $\jset_{\nearrh}$, $\mset_{\nearrh}$) and then retrieves the $\rh$ $\idsender_{\curk}$ whose node identifier is the largest and is greater than or equal to $\idsender_{\curp}$. Upon retrieving $\rh$ $\idsender_{\curk}$, $\ffindpath$ retrieves the $\maxamount_{\nearrh, \curk}$ from the list $\mathcal{M}_{\curk}$($\maxamount_{\nearrh,\curk}$, $\cdot$, $\cdot$, $\cdot$, $\cdot$ ) such that $\maxamount_{\nearrh, \curk}$ $\geq$ $\amount$. $\ffindpath$ retrieves the finger table $\jset_{\curk}$ from the $\rtable$ ($\idsender_{\curk}$, $\jset_{\curk}$, $\mset_{\curk}$) and checks if $\idsender_{\curp}$ $\in$ $\jset_{\curk}$ if yes, $\ffindpath$ retrieves ($\maxamount_{\curk,\curp}$) from $\mset_{\curk}$ ($\maxamount_{\curk,\curp}$, $\cdot$, $\cdot$, $\cdot$, $\cdot$) and checks if $\maxamount_{\curk, \curp}$ $\geq$ $\amount$. If yes, $\ffindpath$ retrieves ($\maxamount_{\curk,\curp}$, $\sigma_{\maxamount_{\curk,\curp}}^{\curk}$, $\sigma_{\maxamount_{\curk,\curp}}^{\curp}$, $tc_{\curk, \curp}$, $tv_{\curk, \curp}$) from $\mset_{\curk}$ present in $\rtable$ ($\idsender_{\curk}$, $\jset_{\curk}$, $\mset_{\curk}$). $\ffindpath$ constructs a tuple $\mathbb{Q}_{\curp}$ and adds ($\idsender_{\curk}$, $\idsender_{\curp}$, $\sigma_{\maxamount_{\curk,\curp}}^{\curk}$, $\sigma_{\maxamount_{\curk,\curp}}^{\curp}$, $tc_{\curk, \curp}$, $tv_{\curk, \curp}$) and samples a $\pid_{\curk}$ $\sample$ $\{0,1\}^\lambda$ and adds the tuple ($\txid$, $\pid_{\curp}$, $\mathbb{Q}_{\curp}$) to the stack $\mathbb{P}$. If no, $\ffindpath$ pops the next element from the stack $\mathbb{W}$. If $\idsender_{\curp}$ $\notin$ $\jset_{\curk}$, $\ffindpath$ retrieves the entry with the node identifier largest and $\geq$ to the node identifier of $\idsender_{\curp}$ and assigns it to $\idsender_{\curk}$.
Once the stack $\mathbb{W}$ is $\emptyset$, (all the $\rh$ have been popped and processed), the $\ffindpath$ updates the $\txtable$ to store ($\idsender_{Alice}$, $\cdot$, $\cdot$, $\cdot$, $\cdot$, $\pset$, $\cdot$, $\cdot$, $\cdot$) and constructs a tuple ($\mathsf{All~Paths}$, $\txid$, $\mathbb{P}$) and sends this tuple to $\nearrh$ and Alice or $\simulator$. 
\end{itemize}
\end{minipage}} 
\end{figure}
\noindent
5) \underline{{$\fhtlc$ functionality \ref{fig:fhtlc}}}: This functionality is depicted in Figure \ref{fig:fhtlc} and handles the HTLC creation and fulfillment functions. This functionality performs two operations. \textbf{Initialization} and \textbf{HTLC fulfillment}. The initialization phase handles the creation of HTLC between the $\send$ and $\rec$. In this phase, the identities of $\send$ and $\rec$ are authenticated and a message is returned to both of them upon successful authentication. The second operation performed by  this functionality is the HTLC fulfillment, where every preceding node on the path communicates the digest and the succeeding node communicates the preimage used to produce the digest. If the hash of the preimage successfully produces the digest, the payment is fulfilled. \\
6) \underline{{$\faux$ functionality}}: This functionality handles the pathfinding between the $\send$, $\nearrh$ and the $\rec$, $\erh$. This functionality internally uses a max-flow algorithm to compute the path from $\send$ to $\nearrh$ and $\erh$ to $\rec$ that has the least hop-count. \\ 
We now give the proof of Theorem\ref{thm:uctheorem}.
\begin{proof}
\label{prf:ucproof}
\textbf{Initialization}: The simulator $\simulator$ simulates the actions of a set of honest nodes $\hset$ $\subset$ $[1 .. n]$ and the adversary $\adversary$ simulates the actions of a set of dishonest nodes, $\dset$ $\subset$ $[1 .. n]$. 
For each node $i$ $\in$ $\hset$, $\simulator$ generates the input tuples ($\mathsf{Key \ Gen}$, $\idsender_{i}$) and sends them to the $\finit$ functionality. The $\finit$ functionality internally calls the $\fsigverify$ functionality and forwards the tuple ($\mathsf{Key \ Gen}$, $\idsender_{i}$) sent by the $\simulator$ to the $\fsigverify$ functionality. 
For each node $i$ $\in$ $\hset$, the $\finit$ sends a tuple ($\mathsf{Long \ Term \ Verification \ Key}, \idsender_{i}, \vk_{i}$) and
($\mathsf{Temporary \ Verification \ Key},  \idsender_{i}, \VK_{i}$) to $\simulator$. For each node $i$ $\in$ $\hset$, $\simulator$, generates the input ($\sign$, $\idsender_{i}$, $\VK_{i}$) and sends it to $\finit$ functionality. $\finit$ internally calls $\fsigverify$ and forwards the tuple ($\sign$, $\idsender_{i}$, $\VK_{i}$) sent by $\simulator$. Upon receiving the tuple ($\signature$, $\idsender_{i}$, $\VK_{i}$, $\sigma_{i}$), from $\fsigverify$, this tuple is forwarded to $\simulator$ by the $\finit$ functionality. The adversary $\adversary$ generates the tuples ($\mathsf{Long \ Term \ Verification \ Key}, \idsender_{j}, \vk_{j}$) and ($\mathsf{Temporary \  Verification \ Key}, \idsender_{j}, \VK_{j}$), for those nodes $j$ $\in$ $\dset$ that have a direct connection with the nodes $\in$ $\hset$ and gives them to the simulator $\simulator$. The adversary $\adversary$ also generates the tuple ($\mathsf{Verify}$, $\VK_{j}$, $\sigma_{\VK_{j}}$, $\vk_{j}$) for those nodes $j$ $\in$ $\dset$ that have a direct connection with the nodes $\in$ $\hset$ and sends it to $\simulator$.
The simulator $\simulator$ generates the inputs ($\mathsf{Immediate \ Neighbors}$, $\iset_{i}$) for each node $i$ $\in$ $\hset$ and sends this tuple to the $\finit$ functionality. 
\begin{figure}[H]
\caption{$\fhtlc$ ideal functionality} 
\label{fig:fhtlc}
\fbox{\begin{minipage}{37em}
\flushleft
\begin{itemize}
\item \textbf{Initialization:} Upon receiving the tuple ($\mathrm{Payment}$, $\vk_{Bob}$, $\vk_{Alice}$, $\idsender_{Alice}$) from $\fpayment$, $ \fhtlc$ checks if $\vk_{Alice}$, $\vk_{Bob}$ $\in$ $\utable$ = ($\idsender_{Alice}$, $\vk_{Alice}$ $\cdot$, $\cdot$, $\cdot$) and $\utable$ = ($\idsender_{Bob}$, $vk_{Bob}$, $\cdot$, $\cdot$, $\cdot$) respectively. If yes, $\fhtlc$ sends a message ($\mathrm{Init \ OK}$) to Bob. Else it returns a $\bot$. Upon receiving the tuple ($\mathrm{Payment}$, $\vk_{Bob}$, $\vk_{Alice}$, $\idsender_{Alice}$) from $\fpayment$, $ \fhtlc$ checks if $\vk_{Alice}$, $\vk_{Bob}$ $\in$ $\utable$ = ($\idsender_{Alice}$, $\vk_{Alice}$ $\cdot$, $\cdot$, $\cdot$) and $\utable$ = ($\idsender_{Bob}$, $vk_{Bob}$, $\cdot$, $\cdot$, $\cdot$) respectively. If yes, $\fhtlc$ sends a message ($\mathrm{Init \ OK}$) to Alice.  Else it returns a $\bot$. 
\item \textbf{HTLC fulfillment:} Upon receiving the tuple ($\mathrm{HTLC \ tuple}$, $\vk_{i}$, $\vk_{i+1}$, $\txid$, $\amount$, $Y$, $X$)   for every pair of consecutive nodes along the path of the $\txid$ from Bob to Alice, from $\fpayment$ functionality, $\fhtlc$ checks if $H(X)=Y$, if yes, $\fhtlc$ sends a message ($\mathrm{Success}$) to $\fpayment$. Else it returns a $\bot$.
\end{itemize}
\end{minipage}}
\end{figure}
$\simulator$ generates the input tuple ($c_{\langle \curi, \curk \rangle}$, $v, t, f$) for each node $k$ $\in$ $\jset_{i}$ and sends this tuple to the $\fdht$. $\fdht$ internally calls the $\fpcn$ functionality and forwards the tuple ($c_{\langle \curi, \curk \rangle}$, $v, t, f$) to $\fpcn$. Upon receiving the value $ h_{i}$ for each node $i$ $\in$ $\hset$, this value is sent by $\fpcn$ to the $\fdht$ which sends it to $\simulator$. Upon receiving the tuple ($\mathsf{Verify}$, $\VK_{i}$, $\sigma_{\VK_{i}}$, $\vk_{i}$) from $\simulator$, for all the nodes in $\iset_{i}$, $\finit$ functionality forwards this tuple to the $\fsigverify$. $\forall$ nodes $i$, that have a direct connection with the nodes $\in$ $\dset$, $\simulator$ forwards the tuple ($\mathsf{Verify}$, $\VK_{i}$, $\sigma_{\VK_{i}}$, $\vk_{i}$) to $\adversary$. 
\begin{figure}[H]
\caption{$\fpcn$ ideal functionality \cite{malavolta2017concurrency}}
\fbox{\begin{minipage}{37em}
\flushleft
\textbf{Open Channel}: 
On input ($\mathsf{open}$, $c_{\langle u, u^\prime \rangle}$, $v, u^\prime, t, f$) from a user u, the $\mathcal{F}$ checks whether $u,u^\prime$ is well-formed (contains valid identifiers and it is not a duplicate) and eventually sends ($c_{\langle u, u^\prime \rangle}, v, t, f$) to $u$, who can either abort or authorize the operation. In the latter case, $\mathcal{F}$ appends the tuple ($c_{\langle u, u^\prime \rangle}, v, t, f$) to $\mathsf{B}$ and the tuple ($c_{\langle u, u^\prime \rangle}, v, t, h$) to $\mathcal{L}$, for some random $h$. $\mathcal{F}$ returns $h$ to $u$ and $u^\prime$. \\
\textbf{Close Channel}: 
On input ($\mathsf{close}$, $c_{\langle u, u^\prime,  \rangle}, h$) from a user $\in$ {$u, u^\prime$} the ideal functionality $\mathcal{F}$ parses $\mathsf{B}$ for an entry ( $c_{\langle u, u^\prime \rangle}$, $v, t, f$) and $\mathcal{L}$ for an entry ( $c_{\langle u, u^\prime \rangle}$, $v^\prime, t^\prime, h$), for h $\neq$ $\bot$. If $c_{\langle u, u^\prime \rangle}$ $\in$ $\mathcal{C}$ or $t$ >|$\mathcal{B}$|or $t^\prime$ $>$ |$\mathcal{B}$| the functionality aborts. Otherwise, $\mathcal{F}$ adds the entry ( $c_{\langle u, u^\prime \rangle}$, $ u^\prime, v^\prime, t^\prime$) to $\mathcal{B}$ and adds $c_{\langle u, u^\prime \rangle}$ to $\mathcal{C}$. $\mathcal{F}$ then notifies both users involved with a message ($c_{\langle u, u^\prime \rangle}, \bot, h $).
\flushleft
\textbf{Payment}: 
On input ($\mathsf{pay}$, $v, (c_{\langle u_{0}, u_{1}\rangle}, \cdots c_{\langle u_{n}, u_{n+1}}), (t_{0}, \cdots, t_{n})$) from a user $u_{0}$, $\fpcn$ executes the following interactive protocol.
    \begin{enumerate}
    \item For all $i$ $\in$ $\{1, \cdots, (n+1)\}$ $\fpcn$ samples a random $h_{i}$ and prases $\mathsf{B}$ for an entry of the form ($c_{\langle u_{i-1}, u_{i}^\prime\rangle}, v_{i}, t_{i}^\prime, f_{i}$). If such an entry does exist $\fpcn$ sends the tuple ($h_{i}, h_{i+1}, c_{\langle u_{i-1}, u_{i}\rangle},c_{\langle u_{i}, u_{i+1}\rangle}$, $v-$  $\sum_{j=1}^{n} f_{j}, t_{i-1},t_{i}$) to the user $u_{i}$ via an anonymous channel (for the specific case of receiver the tuple is only ($h_{n+1}, c_{\langle u_{n}, u_{n+1} \rangle, v, t_{n}}$)). Then $\fpcn$ checks whether for all entries of the form ($c_{\langle u_{i-1},u_{i} \rangle}, v_{i}^\prime, \cdot, \cdot$) $\in$ $\lset$ it holds that $v_{i}^\prime$ $\geq$ $\Biggl(v-\sum_{j=1}^{n}f_{j}\Biggl)$ and that $t_{i-1}$ $\geq$ $t_{i}$. If this is the case $\fpcn$ adds $d_{i}$= ($c_{\langle u_{i-1},u_{i}},v_{i}^\prime, \cdot,, \cdot$) $\in$ $\lset$ is the entry with the lowest $v_{i}^\prime$. If any of the conditions are not met, $\fpcn$ removes from $\lset$ all entries $d_{i}$ added in this phase and aborts.
    \item For all $i$ $\in$ \{($n+1$), $\cdots, 1$\} $\fpcn$ queries all $u_{i}$ with ($h_{i},h_{i+1}$) through an anonymous channel. Each user can reply with either $T$ or $\bot$. Let $j$ be the index of the user that returns $\bot$ such that for all $i$ $\geq$ $j$ $:$ $u_{i}$ returned $T$. If no user returned $\bot$ we set $j=0$. 
    \item For all $i$ $\in$ \{$j+1, \cdots, n$\} the ideal functionality $\fpcn$ updates $d_{i}$ $\in$ $\lset$ (defined as above) to $(--, --, h_{i})$ and notifies the user of success of the operation with some distinguished message ($\mathsf{success}, h_{i}, h_{i+1}$). For all $i$ $\in$ \{$0, \cdots, j$\} (if $j \neq 0$) $\fpcn$ removes $d_{i}$ from $\lset$ and notifies the user with the message ($\bot, h_{i}, h_{i+1}$).
\end{enumerate}
   \end{minipage}}
\end{figure}
Upon receiving the tuple ($\verification$, $\idsender_{i}$, $\VK_{i}$, $f$) from $\fsigverify$, $\finit$ sends this tuple to the $\simulator$. 
$\forall$ nodes $j$ in $\dset$ that have a direct connection with the nodes $\in$ $\hset$, $\simulator$ will send the tuple ($\verification$, $\idsender_{j}$, $\VK_{j}$, $f$) to the adversary $\adversary$. If the value of $f$ for the node simulated by $\adversary$ $\neq$ 1, the $\fsigverify$ returns a $\bot$ and aborts.  \\
\textbf{DHT Initialization}: For every node $i$ $\in$ $\hset$, $\simulator$ generates the input tuple ($\mathsf{DHT \ Join}$, $\idsender_{i}$) and sends it to the $\fdht$ functionality. The $\fdht$ functionality computes the $l_{k}$ $\forall$ $k$ $\in$ [$1, \cdots, m$] and adds the $l_{k}$ to the list $\jset_{i}$ for each $i$. Upon performing this computation, the $\fdht$ functionality sends the tuple ($\mathsf{FingerTable}$, $\jset_{i}$) to $\simulator$. $\simulator$ will generate the input tuple ($\mathsf{Open}$, $c_{\langle \curi, \curk \rangle}, v, \curi, t, f$) for each node $i$ $\in$ $\hset$ and for each node $k$ $\in$ $\jset_{i}$ and sends this tuple to $\fdht$. $\fdht$ internally calls $\fpcn$ functionality and sends the tuple ($\mathsf{Open}$, $c_{\langle \curi, \curk \rangle}, v, \curi, t, f$) to $\fpcn$. 
\begin{figure}[H]
\caption{$\mathcal{F}_{SIG}$ functionality \cite{canetti2004universally}}
\fbox{\begin{minipage}{37em}
\flushleft
\textbf{Key Generation:} Upon receiving a value ($\mathsf{KeyGen}$, $sid$) from some party $S$, verify that $sid= (S, sid ^\prime)$ for some $sid ^\prime$. If not, then ignore the request. Else, hand ($\mathsf{KeyGen}$,$sid$) to the adversary. Upon receiving ($\mathsf{Verification \ Key}$, $sid$, $v$) from the adversary, output($\mathsf{Verification \ Key}$, $sid, v$) to $S$, and record the pair($S,v$). \\
\textbf{Signature Generation:} Upon receiving a value ($\mathsf{Sign}$, $sid, m$) from $S$, verify that $sid=(S, sid^\prime)$ for some $sid^\prime$. If not, then ignore the request. Else, send ($\mathsf{Sign}$, $sid, m$) to the adversary. Upon receiving ($\mathsf{Signature}$, $sid, m, \sigma$) from the adversary, verify that no entry ($m, \sigma, v, 0$) is recorded. If it is, then output an error message to $S$ and halt. Else,output ($\mathsf{Signature}, sid, m, \sigma$) to $S$, and record the entry ($m, \sigma, v, 1$). \\
\textbf{Signature Verification:} Upon receiving a value ($\mathsf{Verify}$, $sid, m, \sigma, v^\prime$) from some party $P$, hand ($\mathsf{Verify}$, $sid, m, \sigma, v^\prime$)  to the adversary. Upon receiving ($\mathsf{Verified}$, $sid, m, \phi$) from the adversary do:
\begin{enumerate}
\item If $v^\prime$= $v$ and the entry ($m, \sigma, v, 1$) is recorded, then set $f=1$. (This condition guarantees completeness: If the verification key $v^\prime$ is the registered one and $\sigma$ is a legitimately generated signature form, then the verification succeeds.)
\item Else, if $v^\prime=v$, the signer is not corrupted, and no entry ($m, \sigma ^ \prime, v, 1$) for any $\sigma^\prime$ is recorded, then set $f=0$ and record the entry ($m, \sigma, v, 0$). (This condition guarantees unforgeability: If $v^\prime$ is the registered one, the signer is not corrupted, and never signed $m$, then the verification fails.)   
\item Else, if there is an entry ($m, \sigma, v^\prime, f^\prime$) recorded, then let $f=f^\prime$. (This condition guarantees consistency: All verification requests with identical parameters will result in the same answer.)
\item Else, let $f=\phi$ and record the entry ($m,\sigma,v^\prime,\phi$).
\end{enumerate}
Output ($\mathsf{Verified}$, $id, m, f$)to $P$.

\end{minipage}}
    
    \label{fig:my_label}
\end{figure} 
Similarly, $\adversary$ generates the tuple ($\mathsf{Open}$, $c_{\langle \curi, j \rangle}, v, j, t, f$) for every node $j$ $\in$ $\dset$ that has a direct connection with all the  nodes $i$   $\in$ $\jset_{i}$ and sends it to $\simulator$. $\simulator$ forwards this tuple to the $\fdht$ functionality. If the node $j$ $\notin$ $\jset_{\curi}$, $\fdht$ returns a $\bot$ and aborts. Else, $\fdht$ internally calls the $\fpcn$ functionality and forwards the tuple ($\mathsf{Open}$, $c_{\langle \curi, j \rangle}, v, j, t, f$) to $\fpcn$. $\adversary$ also generates the input tuple ($c_{\langle \curi, j \rangle}$, $v, t, f$) for every node $j$ $\in$ $\dset$ that has a direct connection with all the  nodes $i$ $\in$ $\jset_{i}$ and sends to $\simulator$. $\simulator$ forwards this tuple to the $\fdht$ functionality who in turn forwards this tuple to the $\fpcn$ functionality. Let us consider four cases:

\begin{enumerate}
\item \textbf{Case 1: $\rh$ $\curi$ $\in$ $\hset$, $\curk$ $\in$ $\jset_{\curi}$ $\in$ $\hset$}: $\simulator$ generates the input tuples ($\mathsf{Maximum \ Amount}$, $\maxamount_{i,k}$, $\sigma_{\maxamount_{i,k}}^i$) and ($\maxamount_{i,k}$, $\sigma_{\maxamount_{i,k}}^k$) for each node $\curi$ $\in$ $\hset$ and for each node $ \curk$ $\in$ $\jset_{i}$ and sends these tuples to the $\fdht$ functionality. Once the time epoch $\delta$ chosen by the $\fdht$ functionality expires, $\simulator$ generates the input tuples ($\mathsf{Maximum~Amount} ^ \prime$, $\maxamount_{i,k} ^ \prime$, $\sigma_{\maxamount_{i,k} ^ \prime}^{i}$) and ($\mathsf{Maximum~Amount} ^ \prime$, $\maxamount_{i,k} ^ \prime$, $\sigma_{\maxamount_{i,k} ^ \prime}^{k}$) and sends them to the $\fdht$ functionality. Upon receiving the input tuples from $\simulator$, the $\fdht$ functionality adds the of $tc$ $\leftarrow$ $\tset$ and $tv$ $\leftarrow$ $\tset$ to the tuples ($\mathsf{Maximum \ Amount}$, $\maxamount_{i,k}$, $\sigma_{\maxamount_{k,i}}^i$, $tc_{i,k}$, $tv_{i,k}$), ($\mathsf{Maximum \ Amount}$, $\maxamount_{i,k}$, $\sigma_{\maxamount_{k,i}}^k$, $tc_{i,k}$, $tv_{i,k}$) or adds the values $tc_{i,k}^\prime$ $\leftarrow$ $\tset$ and $tv_{i,k}^\prime$ $\leftarrow$ $\tset$ to the tuples ($\mathsf{Maximum~Amount} ^ \prime$, $\maxamount_{i,k} ^ \prime$, $\sigma_{\maxamount_{i,k} ^ \prime}^{i}$, $t_{c_{i,k}}^\prime$, $t_{v_{i,k}}\prime$) and ($\mathsf{Maximum~Amount} ^ \prime$, $\maxamount_{i,k} ^ \prime$, $\sigma_{\maxamount_{i,k} ^ \prime}^{i}$, $t_{c_{i,k}}, t_{v_{i,k}}$).

\begin{figure}[H]
\caption{$\faux$ Ideal Functionality}
\fbox{\begin{minipage}{37em}
\flushleft
\begin{itemize}
\item \textbf{Sender to $\nearrh$ path computation}: Upon receiving the tuple ($\idsender_{Alice}$, $\amount$, $\sigma_{\amount}$, $\txid$) from Alice or $\simulator$, $\faux$ retrieves the values of $f$ and $\VKalice$ from the $\utable$ ($\idsender_{Alice}$, $\cdot$, $\VKalice$, $\cdot$, $f$). If the value of $f=0$ or if $f=\phi$, $\faux$ returns a $\bot$ and aborts. If the value of $f=1$, $\faux$ sends the tuple ($\verification$, $\idsender_{Alice}$, $\amount$, $\sigma_{\amount}$, $\VKalice$) to $\fsigverify$.  Upon receiving the tuple ($\mathsf{Verified}$, $\idsender_{Alice}$, $\amount$, $\fprime$), from $\fsigverify$, $\faux$ checks the value of $\fprime$. If $\fprime$ = 0 or if $\fprime=\phi$, $\faux$ returns a $\bot$ and aborts. Else $\faux$ internally runs a max-flow algorithm and computes the $\rh$ closest to Alice in terms of hop-count that satisfies the $\amount$ that Alice intends to route. $\faux$ creates the tuple ($\VK_{1}$, $\dots$, $\VK_{m}$) which are the nodes along the path from Alice to $\nearrh$, where $m \in \mathbb{U}$. $\faux$ creates a tuple  ($\nearrh$, $\idsender_{\nearrh}$, $\mathsf{Path}$, ($\VK_{1}$, $\dots$, $\VK_{m}$)) and sends this tuple to Alice and $\simulator$ and adds the nodes ($\VK_{1}$, $\dots$, $\VK_{m}$) to a set $\mathbb{O}$ that it locally maintains and updates the $\txtable$ to store ($\idsender_{Alice}$, $\cdot$, $\mathbb{O}$, $\cdot$, $\cdot$, $\cdot$, $\cdot$, $\cdot$, $\cdot$, $\cdot$) 
\end{itemize}
\flushleft
\begin{itemize}
\item \textbf{Receiver to $\erh$ path computation}: Upon receiving the tuple ($\idsender_{Bob}$, $\idsender_{\erh_{i}}$) from Bob or $\simulator$ for $i$ $\in$ $\mathbb{R}^\prime$, where $\mathbb{R}^\prime$ $\subset$ $\mathbb{RH}$. $\faux$ retrieves the $\VK_{Bob}$ from the $\utable$ ($\idsender_{Bob}$, $\cdot$, $\VK_{Bob}$, $\cdot$, $f$). If the value of $f=0$ or if the value of $f=\phi$, $\faux$ returns a $\bot$ and aborts. If the value of $f=1$, $\faux$ internally uses a maximum flow algorithm to compute the paths in terms of hop-count that can satisfy the $\amount$ that Alice intends to route to all the $\rh$ $i$ $\in$ $\mathbb{R}^\prime$, and creates a tuple ($\erh_{i}$, $\idsender_{\erh_{i}}$,$\mathsf{Path}$, ($\VK_{1}^\prime$, $\dots$, $\VK_{m}^\prime$)), where $m$ $\in$ $\mathbb{U}$ and sends it to Bob or $\simulator$ and adds the nodes ($\VK_{1}^\prime$, $\dots$, $\VK_{m}^\prime$) to a set $\mathbb{O}^\prime$ that it locally maintains and updates the $\txtable$ to store ($\idsender_{Alice}$, $\cdot$, $\cdot$, $\mathbb{O}^\prime$, $\cdot$, $\cdot$, $\cdot$, $\cdot$, $\cdot$, $\cdot$,$\cdot$).


\end{itemize}
\end{minipage}} 
\end{figure}

\item \textbf{Case 2: $\rh$ $\curi$ $\in$ $\hset$, $\curk$ $\in$ $\jset_{\curi}$ $\in$ $\dset$}: 
$\simulator$ generates the input tuples ($\mathsf{Maximum \ Amount}$, $\maxamount_{i,k}$, $\sigma_{\maxamount_{i,k}}^i$) for each node $\curi$ $\in$ $\hset$. Similarly, $\adversary$ generates ($\mathsf{Maximum \ Amount}$, $\maxamount_{i,j}$, $\sigma_{\maxamount_{i,j}}^j$) for each node $j$ $\in$ $\dset$ and $\in$ $\jset_{\curi}$ and sends it to $\simulator$. Once the time epoch $\delta$ chosen by  the $\fdht$ functionality expires, $\simulator$ generates the input tuples ($\mathsf{Maximum~Amount} ^ \prime$, $\maxamount_{i,k} ^ \prime$, $\sigma_{\maxamount_{i,k} ^ \prime}^{i}$). Once the time epoch $\delta$ has expired, $\adversary$ constructs the tuple ($\mathsf{Maximum~Amount} ^ \prime$, $\maxamount_{i,k} ^ \prime$, $\sigma_{\maxamount_{i,k} ^ \prime}^{k}$) and sends it to $\simulator$. $\simulator$ forwards the input tuples generated by itself and by  $\adversary$ to the $\fdht$ functionality. Upon receiving these tuples, for every node $j$, being simulated by $\adversary$, $\fdht$ functionality checks if $j$ $\in$ $\jset_{\curi}$, if not, it returns a $\bot$ and aborts.


\item \textbf{Case 3: $\rh$ $\curi$ $\in$ $\dset$, $\curk$ $\in$ $\jset_{\curi}$ $\in$ $\hset$}:
$\adversary$ creates the input tuple ($\mathsf{Maximum \ Amount}$, $\maxamount_{i,k}$, $\sigma_{\maxamount_{i,k}}^i$) for the node $\curi$ $\in$ $\dset$ and send this tuple to $\simulator$. If the epoch $\delta$ chosen expires, $\adversary$ constructs the input tuple ($\mathsf{Maximum~Amount} ^ \prime$, $\maxamount_{i,k} ^ \prime$, $\sigma_{\maxamount_{i,k} ^ \prime}^{i}$) and sends it to $\simulator$. Upon receiving the input tuple from $\adversary$, $\simulator$ constructs the input tuple ($\mathsf{Maximum \ Amount}$, $\maxamount_{i,k}$, $\sigma_{\maxamount_{i,k}}^k$) for each node $\curk$ $\in$ $\jset_{\curi}$. If the time epoch $\delta$ expires, $\simulator$ will construct the input tuple \\ ($\mathsf{Maximum~Amount} ^ \prime$ , $\maxamount_{i,k} ^ \prime$, $\sigma_{\maxamount_{i,k} ^ \prime}^{k}$). $\simulator$ will send the tuples it constructed along with the tuple sent by $\adversary$ to the $\fdht$ functionality. Upon receiving these tuples, $\fdht$ checks if $\curi$ $\in$ $\rtable$, if not, the $\fdht$ functionality returns a $\bot$ and aborts.
\item \textbf{Case 4: $\rh$ $\curi$ $\in$ $\dset$, $\curk$ $\in$ $\jset_{\curi}$ $\in$ $\dset$}: 
This is handled locally by $\adversary$ and we do not simulate it.
\end{enumerate}


For a node $i$ $\in$ $\hset$ leaving the DHT ring, the simulator $\simulator$ constructs an input tuple ($\mathsf{DHT~Leave}$, $\idsender_{\curi}$)and sends it to the $\fdht$ functionality. Upon receiving the tuple from $\simulator$, the $\fdht$ functionality deletes the entry ($\idsender_{\curi}$, $\jset_{k}$, $\mset_{\curi}$) from the $\rtable$. The case of $\curi$ $\in$ $\dset$ is handled locally by the adversary $\adversary$. \\

\par 
\noindent
\textbf{PathFinding between $\send$ and $\nearrh$, $\erh$ and $\rec$}: The next phase is pathfinding between the $\send$ and $\nearrh$ and between the $\erh$ and the $\rec$ and we have two cases to consider here \\ 
\begin{enumerate}

\item \textbf{Case 1:} $\send$ $\in$ $\hset$: The simulator $\simulator$ constructs an input tuple  ($\idsender_{\send}$, $\amount$, $\sigma_{\send_{\amount}}$, $\txid$) and sends this tuple to the $\faux$ functionality. The pathfinding between the $\send$ and the $\nearrh$ is done by $\faux$ functionality. This functionality returns the tuple ($\nearrh$, $\idsender_{\nearrh}$, $\mathsf{Path}$, $\VK_{1}, \cdots, \VK_{m}$) to $\simulator$.  
\item \textbf{Case 2:} $\rec$ $\in$ $\hset$: $\simulator$ constructs an input tuple  ($\idsender_{\rec}$, $\idsender_{\erh_{i}}$) for $\curi \in \mathbb{R}^\prime$, where $\mathbb{R}^\prime$ $\subset$ $\mathbb{RH}$ and sends this tuple to the $\faux$ functionality. The pathfinding between the $\rec$ and $\erh$ $\curi$ is done by $\faux$ functionality.
For each $\erh$ $\curi$, this functionality returns the tuple ($\curi$, $\idsender_{\curi}$, $\mathsf{Path}$, ($\VK_{1}^\prime$, $\cdots$, $\VK_{m}^\prime$)) to $\simulator$.  
\end{enumerate}


  \par 
  \noindent
  \textbf{PathFinding inside the DHT ring}: The next phase is pathfinding between the $\nearrh$ and the other nodes in the DHT ring and we have four cases here to consider here.

  \begin{enumerate}

\item \textbf{Case 1:} $\send$ and the $\nearrh$ $\in$ $\hset$:
 $\simulator$ constructs the tuple ($\verify$, $\idsender_{\send}$, $\amount$, $\sigma_{\amount_{i}}$, $\VK_{i}$) for a $\send$ $\in$ $\hset$ and sends this tuple to the $\fsigverify$ functionality. Upon receiving the tuple ($\mathsf{Verified}$, $\idsender_{i}$, $\amount$, $f$) from the $\fsigverify$ functionality, $\simulator$ constructs the input tuple ($\txid$, $\mathsf{Verified}$, $\idsender_{\curi}$, $\amount$, $f$, $\idsender_{\nearrh}$), where $\nearrh$ $\in$ $\hset$, and sends this tuple to the $\ffindpath$ functionality. The $\ffindpath$ return the tuple  ($\mathsf{All~Paths}$, $\txid$, $\pset$) to $\simulator$.

\item \textbf{Case 2:} $\send$ $\in$ $\hset$ and $\nearrh$ $\in$ $\dset$: $\simulator$ constructs the tuple ($\mathsf{Verify}$, $\idsender_{\send}$, $\amount$, $\sigma_{\amount_{\send}}$, $\VK_{\send}$) for a $\send$ $\in$ $\hset$ and sends this tuple to the $\fsigverify$ functionality. $\simulator$ sends $\txid$ to $\adversary$. $\adversary$ sends $\idsender_{\nearrh}$ to  $\simulator$. $\simulator$ forwards the $\idsender_{\nearrh}$ to the $\fdht$ functionality. If the $\idsender_{\nearrh}$ $\notin$ $\rtable$, the functionality returns a $\bot$ and aborts. Else, the functionality returns the $\idsender_{\nearrh}$ to $\simulator$.
The adversary $\adversary$ will send the tuple ($\mathsf{All~Paths}$, $\txid$, $\pset$) to $\simulator$. $\simulator$ forwards this tuple to the $\fdht$ functionality. The $\fdht$ functionality validates this tuple and aborts if any of the entries in this tuple are incorrect. 
\item \textbf{Case 3:} $\send$ $\in$ $\dset$ and $\nearrh$ $\in$ $\hset$: 
The $\adversary$ does $\txid$ $\sample$ $\{0,1\}^\lambda$ and constructs an input tuple ($\mathsf{Verify}$, $\idsender_{\send}$, $\amount$, $\sigma_{\amount_{\send}}$, $\VK_{\send}$) and this tuple is sent to $\simulator$.  Upon receiving this tuple from $\adversary$, $\simulator$ forwards this tuple to the $\fsigverify$ functionality. Upon receiving the tuple ($\mathsf{Verified}$, $\idsender_{\send}$, $\amount$, $\sigma_{\amount_{\send}}$, $f$) from $\fsigverify$, $\simulator$ constructs the tuple ($\txid$, $\mathsf{Verified}$, $\idsender_{\curi}$, $\amount$, $f$,$\idsender_{\nearrh}$) for a $\nearrh$ $\in$ $\hset$ and sends it to the $\ffindpath$ functionality. If the value of $f$ $\neq$ 1, the functionality returns a $\bot$ and aborts.
Upon receiving the tuple ($\mathsf{All~Paths}$, $\txid$, $\pset$) from the $\ffindpath$ $\simulator$ sends this tuple to $\adversary$. 
\item \textbf{Case 4:} $\send$ $\in$ $\dset$ and $\nearrh$ $\in$ $\dset$: This case is handled locally by $\adversary$. 
\end{enumerate}
\par 
\noindent
\textbf{Payment Phase}: The next phase is the payment. We now consider the possible cases

\begin{enumerate}

\item \textbf{Case 1:} $\send$, $\rec$ and intermediate nodes $\in$ $\hset$: $\simulator$ constructs a tuple ($\mathsf{list~of~\erh}$, $\tset$, $\txid$, $\VK_{\send}$) to the $\fpayment$ and the $\fpayment$ sends the tuple back to the $\simulator$ (playing the role of $\rec$). The $\simulator$ constructs an input tuple ($\erh$, $\erh_{\rec}$, $\txid$, $\VK_{\rec}$, $Y$, $X$) to the $\fpayment$ functionality. The functionality sends this tuple back to $\simulator$ (simulating the $\send$). $\simulator$ (simulating the intermediate nodes) constructs the tuple ($\mathsf{Final~Path}$, $\VK_{i}$, $\VK_{i+1}$) for every pair of consecutive nodes along the path from the $\send$ to the $\rec$ and sends this tuple to the $\fpayment$ functionality. The $\fpayment$ sends this tuple back to $\simulator$. Upon receiving the tuple ($\mathsf{Final~Path}$, $\VK_{i}$, $\VK_{i+1}$) for every pair of consecutive nodes along the path, $\simulator$ constructs a tuple ($\mathsf{HTLC~Tuple}$, $\vk_{i}, \vk_{i+1}$, $\txid$, $\amount$, $Y$, $X$) for every pair of consecutive nodes along the path from the $\send$ to the $\rec$ and sends this tuple to $\fpayment$. $\fpayment$ forwards this tuple to the $\fhtlc$ functionality. $\fpayment$ returns the tuple ($\mathsf{Success}$) to $\simulator$. 

\item \textbf{Case 2:} $\send$, $\rec$ $\in$ $\hset$ and some intermediate nodes $\in$ $\dset$: $\simulator$ (simulating the $\send$) constructs a tuple ($\mathsf{list~of~\erh}$, $\tset$, $\txid$, $\VK_{\send}$) to the $\fpayment$ and the $\fpayment$ sends the tuple back to the $\simulator$ (simulating the $\rec$). The $\simulator$ (simulating the $\rec$) constructs an input tuple ($\erh$, $\erh_{\rec}$, $\txid$, $\VK_{\rec}$, $Y$, $X$)  to the $\fpayment$ functionality. The functionality sends the tuple ($\erh$, $\erh_{\rec}$, $\txid$, $\VK_{\rec}$, $Y$) to   $\simulator$ (simulating the $\send$). In this case, we consider some intermediate nodes to be honest and some to be dishonest (these intermediate nodes could include the routing helpers as well). For all pairs of consecutive nodes $i,i+1$ along the path from $\send$ to $\rec$ such that node $i$ $\in$ $\hset$ and node $i+1$ $\in$ $\dset$, $\simulator$ (simulating the honest node $\curi$) constructs an input tuple ($\mathsf{Final~Path}$, $\VK_{i}$, $\VK_{i+1}$) and sends this tuple to the $\fpayment$ functionality. The $\fpayment$ functionality sends a tuple ($\mathsf{Final~Path}$, $\txid$, $\VK_{i}$, $\VK_{i+1}$) to the $\simulator$ or $\fpayment$ functionality returns a $\bot$ and aborts. The response sent by the $\fpayment$ is sent to $\adversary$ by  $\simulator$. Similarly, for all pairs of consecutive nodes $i,i+1$ along the path from $\send$ to $\rec$ such that node $i$ $\in$ $\dset$ and node $i+1$ $\in$ $\hset$, $\simulator$ sends $\VK_{\curi+1}$ to $\adversary$. $\adversary$ constructs an input tuple ($\mathsf{Final~Path}$, $\txid$, $\VK_{i}$, $\VK_{i+1}$) and sends it to $\simulator$. $\simulator$ sends this tuple to $\fpayment$ functionality. The $\fpayment$ functionality sends the tuple ($\mathsf{Final~Path}$, $\txid$, $\VK_{i}$, $\VK_{i+1}$) to $\simulator$ (simulating node $\curi+1$) or the $\fpayment$ returns a $\bot$ and aborts. This response by $\fpayment$ is sent to $\adversary$ by $\simulator$. If both the intermediate nodes $\curi$, $\curi+1$ are dishonest, they are handled locally by $\adversary$. For every pair of consecutive nodes $\curi$, $\curi+1$ along the path from the $\rec$ to the $\send$, such that $\curi$ $\in$ $\hset$ and $\curi+1$ $\in$ $\dset$, $\simulator$ (simulating the honest node $\curi$) sends ($\vk_{\curi}$, $Y$, $X$) to $\adversary$. $\adversary$ handles the HTLC payment locally and sends the tuple ($\mathsf{HTLC~Tuple}$, $\vk_{\curi}$, $\vk_{\curi+1}$, $\txid$, $\amount$,  $Y$, $X$) to $\simulator$. $\simulator$ forwards this tuple to the $\fpayment$ functionality. The $\fpayment$ functionality sends the tuple ($\mathsf{Sucess}$) or $\bot$ to $\simulator$, which is forwarded to $\adversary$. Similarly for every pair of consecutive nodes $\curi$, $\curi+1$ along the path from $\rec$ to $\send$, where $\curi+1$ $\hset$ and $\curi$ $\in$ $\dset$, $\adversary$ (simulating node $\curi$), sends $\vk_{i}$ to $\simulator$. $\simulator$ constructs an input tuple ($\mathsf{HTLC~Tuple}$, $\vk_{\curi}$, $\vk{\curi+1}$, $\txid$, $\amount$, $Y$, $X$) and sends it to $\fpayment$. The $\fpayment$ functionality sends this tuple to $\fhtlc$ or returns a $\bot$ and aborts. If the tuple is sent to $\fhtlc$ by $\fpayment$, $\fhtlc$ returns a $\mathsf{Success}$ or $\bot$ to $\fpayment$ and this is sent to $\simulator$ by $\fpayment$. $\simulator$ forwards this result to $\adversary$.

\item \textbf{Case 3:} $\send$ $\in$ $\hset$, $\rec$ and some intermediate nodes in $\dset$: The $\simulator$ (simulating the $\send$) constructs an input tuple ($\mathsf{list~of~\erh}$, $\tset$, $\txid$, $\VK_{\send}$) and sends the tuple to $\adversary$. $\adversary$ (simulating the $\rec$) constructs an input tuple ($\erh$, $\erh_{\rec}$, $\txid$, $\VK_{\rec}$) to $\simulator$. $\simulator$ sends this tuple to the $\fpayment$ functionality. The $\fpayment$ sends this tuple back to $\simulator$ or returns a $\bot$ (in the case of the $\erh_{\rec}$ $\notin$ $\tset$) and aborts. $\adversary$ sends the ($\VK_{\send+1}$) to $\simulator$. $\simulator$ (simulating the $\send$), constructs an input tuple ($\mathsf{Final~Path}$, $\txid$, $\VK_{\send}$, $\VK_{\send+1}$) and sends it to the $\fpayment$ functionality. The $\fpayment$ functionality can either send this tuple back to $\simulator$ or can return a $\bot$ and abort.  $\simulator$ sends the $\txid$ to $\adversary$. For every pair of consecutive nodes, $\curi, \curi+1$ along the path from the $\send$ to $\rec$, the construction of the $\mathsf{Final~Path}$ is done as described in \textbf{Case 1}. Similarly for every pair of consecutive nodes $\curi$, $\curi+1$ along the path from the $\send$ to the $\rec$, such that $\curi+1$ $\in$ $\hset$ and $\curi$ $\in$ $\dset$, the construction of the $\mathsf{Final~Path}$ tuples is done as described in \textbf{Case 2}.

For every pair of consecutive nodes $\curi$, $\curi+1$ along the path from the $\rec$ to the $\send$, such that $\curi$ $\in$ $\hset$ and $\curi+1$ $\in$ $\dset$, $\simulator$ (simulating the honest node $\curi$) sends ($\mathsf{Digest}$, $Y$) to every node $\in$ $\hset$ and also to $\adversary$. $\adversary$ handles the HTLC payment locally and sends the tuple ($\mathsf{HTLC~Tuple}$, $\vk_{\curi}$, $\vk_{\curi+1}$, $\txid$, $\amount$,  $Y$, $X$) to $\simulator$. $\simulator$ forwards this tuple to the $\fpayment$ functionality. The $\fpayment$ functionality sends the tuple ($\mathsf{Sucess}$) or $\bot$ to $\simulator$, which is forwarded to $\adversary$. Similarly
for every pair of consecutive nodes $\curi$, $\curi+1$ along the path from $\rec$ to $\send$, where $\curi+1$ $\in$ $\hset$ and $\curi$ $\in$ $\dset$,  $\simulator$ constructs an input tuple ($\mathsf{HTLC~Tuple}$, $\vk_{\curi}$, $\vk{\curi+1}$, $\txid$, $\amount$, $Y$, $X$) and sends it to $\fpayment$. The $\fpayment$ functionality sends this tuple to $\fhtlc$ or returns a $\bot$ and aborts. If the tuple is sent to $\fhtlc$ by $\fpayment$, $\fhtlc$ returns a $\mathsf{Success}$ or $\bot$ to $\fpayment$ and this is sent to $\simulator$ by $\fpayment$. $\simulator$ forwards this result along with the preimage $X$ and the digest $Y$ to $\adversary$. 

\item  \textbf{Case 4:} $\rec$ and some intermediate nodes $\in$ $\hset$, $\send$ and some intermediate nodes $\in$ $\dset$: $\adversary$ (simulating the $\send$), constructs an input tuple ($\mathsf{list~of~\erh}$, $\tset$, $\txid$, $\VK_{\send}$) and sends it to $\simulator$. The $\simulator$ send this to $\fpayment$ functionality. The $\fpayment$ functionality can either send this tuple back to $\simulator$ (simulating the $\rec$) or can return a $\bot$ and abort (in case of the nodes in $\tset$ being incorrect). If this tuple is sent back to $\simulator$, $\simulator$ (simulating the $\rec$), constructs a tuple ($\mathsf{\erh}$, $\erh_{\rec}$, $\txid$, $\VK_{\rec}$,$Y$, $X$) and sends to $\adversary$. For every pair of consecutive nodes, $\curi,\curi+1$ along the path from the $\send$ to the $\rec$, where $\curi$ $\in$ $\hset$ and $\curi+1$ $\in$ $\dset$ the $\mathsf{Final~Path}$ tuple construction happens as described in \textbf{Case 1}. For every pair of consecutive nodes, $\curi,\curi+1$ along the path from the $\send$ to the $\rec$, where $\curi$ $\in$ $\dset$ and $\curi+1$ $\in$ $\hset$ the $\mathsf{Final~Path}$ tuple construction happens as described in \textbf{Case 2}. For every pair of consecutive nodes $\curi$, $\curi+1$ along the path from the $\rec$ to the $\send$, such that $\curi$ $\in$ $\hset$ and $\curi+1$ $\in$ $\dset$,  $\adversary$ sends the tuple ($\mathsf{HTLC~Tuple}$, $\vk_{\curi}$, $\vk_{\curi+1}$, $\txid$, $\amount$,  $Y$, $X$) to $\simulator$. $\simulator$ forwards this tuple to the $\fpayment$ functionality. The rest of the simulation proceeds as in \textbf{Case 3}

 \item  \textbf{Case 5}: $\send$, $\rec$, and all the intermediate nodes $\in$ $\dset$: This case is handelled locally by $\adversary$. 

\item \textbf{Case 6:} $\send$, $\rec$ are $\dishonest$ and some intermediate nodes are $\honest$: $\adversary$, simulating the $\send$ and constructs the tuple containing the list of $\erh$ $\tset$. $ \adversary$, simulating the $\rec$ is also responsible for the selection of one $\erh$. Once this happens, for every pair of consecutive nodes $\curi$, $\curi+1$ along the path from the $\send$ to the $\rec$, where $\curi$ $\in$ $\hset$ and $\curi+1$ $\in$ $\dset$ or  $\curi$ $\in$ $\dset$ and $\curi+1$ $\in$ $\hset$, the construction of $\mathsf{Final~Path}$ tuples proceeds as described in \textbf{Case 2}. During the construction and validation of the $\mathsf{Final~Path}$ tuples, the authenticity of the $\erh$ picked by $\adversary$ is validated. For every pair of consecutive nodes $\curi$, $\curi+1$ along the path from the $\rec$ to $\send$, where $\curi$ $\in$ $\hset$ and $\curi+1$ $\in$ $\dset$ or  $\curi$ $\in$ $\dset$ and $\curi+1$ $\in$ $\hset$, the payment using HTLC proceeds as described in \textbf{Case 3}.
\end{enumerate}

\end{proof}

%% file: conc_full.tex
\section{Conclusion}
\label{sec:conc}
In this paper, we have designed $\ripd$, a PCN pathfinding and routing protocol that uses distributed hash tables to route transactions in PCNs. Our protocol does not need the presence of a trusted third party, is fully decentralized and can route concurrent transactions. $\ripd$ also ensures the privacy of $\send$ and $\rec$, and atomicity of payments. We have demonstrated the efficiency of raced $\ripd$ by evaluating it on real-world transaction data, and have proven the security of $\ripd$ in the UC framework. The ideas presented in $\ripd$ can potentially be leveraged to decentralized networks in diverse domains such as edge computing and IoT networks.


%% file: appendix_full.tex
\appendix
\section{Extended Related Work}
\label{sec:appendrelated}
The distributed version of Dijkstra's shortest path algorithm presented in~\cite{abdelrahman,aly2013securely,aly2015network} requires each vertex in the graph to reveal the vertices it is not connected to, to an ``algorithm designer'' who runs the algorithm. Revealing this information eventually reveals the entire network topology and cannot be leveraged to PCNs because of privacy violations. 

The solution presented 
in \cite{abdelrahman,aly2013securely,aly2015network} for the minimum cost-flow problem also has the same assumptions as the distributed version of the Dijkstra's shortest path algorithm. 

The minimum mean cycle problem presented in \cite{abdelrahman,aly2013securely,aly2015network} is orthogonal to our work since it focuses on finding cycles in graph which have the least number of edges, whereas $\ripd$ focuses on performing secure routing in PCNs. 
The idea proposed by Abdelrahman \emph{et al.} in \cite{abdelauction} proposes an auction mechanism in which sellers sell the maximum flow that is transmittable through them and the bidders bid for these maximum flows. This idea assumes the total amount of flow transmittable through the network (the network throughput) to be public and also requires a trusted entity called ``control agency'' that oversees the auction. This idea cannot be leveraged to perform secure routing in PCNs. PCNs are distributed networks where having a central root of trust is not possible. Revealing the complete throughput of the network for PCNs is a violation of privacy. 
\par
 DHTs were developed initially to facilitate file-sharing among a set of cooperating peers \cite{chord, rowstron2001pastry,maymounkov2002kademlia,zhao2004tapestry}. DHTs are also being explored for solving routing challenges in MANETs (Mobile Ad-hoc Networks), VANETs (Vehicular Ad-hoc Networks), and for data sharing across IoT (Internet Of Things) devices \cite{alshaikh, zahid2022fault, tran,monfared2023darvan}. However, in the case of PCNs, nodes (peers) do not share data/files but send and receive money. Peer-to-peer routing protocols that use DHT do not take part in any payment channel opening/closing, do not interact with a blockchain, and do not route payments among each other. Finally, in DHTs it suffices if a node is able to locate another node in the network for peer-to-peer communication. However in PCNs, in addition to finding efficient paths between nodes, the paths should also have enough liquidity to route the amount specified by the sender.
\section{Overview of Chord}
\label{sec:appendchord}
 \begin{figure*}[h!]
 \centering
     \includegraphics[width=\columnwidth]{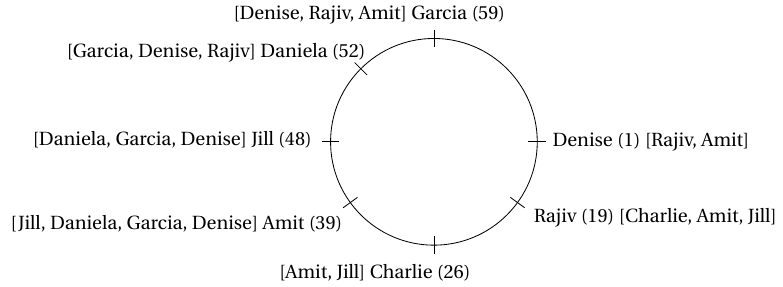}
    \caption{An example Chord ring with 7 routing helpers. The values in parenthesis adjacent to the node represents the node identifier. The values in the square brackets [...] represent finger table entries. In each finger table, we only show unique entries.}
     \label{fig:figchord}
 \end{figure*}
\label{sec:chord}
Chord~\cite {chord} is a scalable, peer-to-peer, distributed lookup protocol that locates a node that stores a particular data item in p2p networks. It uses a consistent hashing mechanism that enables the lookup to be completed in time that is  logarithmic in the number of nodes present in the DHT ring.
The nodes in Chord are placed in the form of a circle called the identifier circle. Each node hashes its IP address to produce an $m$ bit digest that acts as its node identifier, denoted by the numbers next to each node in Figure~\ref{fig:figchord}. 
Each node in the Chord ring in Figure \ref{fig:figchord} is responsible for storing a key (represented as a digest) that points to a certain fragment of data. This key, $\key$ is the digest obtained by hashing the identifier of the key with the same hash function that was used to create the node identifiers.
Each key $\key$ will be assigned to the node whose identifier is equal to or follows the identifier of $\key$ in the identifier circle. Each node in the Chord ring maintains a look-up table called \emph{finger table} that contains at most $m$ entries with $log(m)$ being distinct. Each node also maintains a table containing its first $log(n)$ successors, called the successor table. The first entry in a node's finger table is the node's immediate successor in the identifier circle.        
Consider Figure~\ref{fig:figchord}, where seven nodes are a part of a Chord ring. The finger table entries of a node identified by $i$ are computed thus: (${i}$ + $2 ^ {(i-1)}$ $\mod$ $m$). If we set $m=6$, the finger table entries of node Charlie are: ($26+2^0 \mod 2^ 6$), ($26+2^1 \mod 2 ^ 6$), ($26+2^2 \mod 2 ^ 6$), ($26+2^3 \mod 2 ^ 6$), ($26+2^4 \mod 2 ^ 6$), ($26+2^5 \mod 2 ^ 6$) which gives us the set of identifiers \{27, 28, 30, 34, 38, 61\}, which map to nodes [Amit, Amit, Amit, Amit, Jill and Garcia]. In case the identifier is not assigned to any node in the Chord ring (27 in this example), the corresponding finger table entry would be the next node in the ring whose identifier is greater than 27, in this case, Amit. 
In Chord, 
when a node receives a request to locate a key $\key$ that is not in its possession, it forwards the request to the closest predecessor of $\key$ in its finger table. For example, if Denise wants to resolve a query to locate node Jill, Denise needs to locate the node that precedes Jill in the Chord ring, which is Amit. Now from Denise's finger table shown in Figure \ref{fig:figchord}, the node closest to Amit (based on node identifiers) in Alice's finger table, which contains [Rajiv, Rajiv, Rajiv, Rajiv, Rajiv, Amit] is Amit himself (Amit is in the finge table of Denise), and Jill is the first node in the finger table of Amit which contains [Jill, Daniela, Garcia, Denise]. Hence the distance between Denise and Amit is greater than the distance between Amit and Jill, so Amit is closer to Jill than Denise. Hence Denise passes the request of locating Jill, to Amit. Since the finger table of Amit contains [Jill, Daniela, Garcia, Denise], Jill reaches Amit in one hop. In this manner, the number of steps is halved every time a node locates another node in the identifier circle. This reduces the lookup time to $log(n)$, where $n$ is the number of nodes in the Chord ring.
\section{Protocols}
\label{sec:appendixproto}
 \begin{algorithm}[htbp]
\caption{$\mathsf{Key \ Setup}$}
\label{algo:keysetup}
\For{$i=1;i\leq n;i++$}{
node $i$ does $\mathsf{KeyGen}$($1 ^ \lambda$) $\rightarrow$ $\sk_{i},\vk_{i}$\\
\tcc{creating temporary identities}
node $i$ does $\mathsf{KeyGen}$($1 ^ \lambda$) $\rightarrow$ $\SK_{i},\VK_{i}$\\
node $i$ does  
$\sign_{\sk_{i}}$($\VK_{i}$) $\rightarrow$ $\sigma_{\VK_{i}}$ \\

node $i$ calls $\retrieveneighbors$($\vk_{i}$) $\rightarrow$ $\mathbb{I}_{i}$\\ 
node $i$ sends $\vk_{i}$ to all the nodes in $\mathbb{I}_{i}$\\
\For{j=1;j$\leq$ $|\mathbb{I}_{i}|$;j++}{

 \If{$\verify_{\vk_{i}}$( $\VK_{i}$, $\sigma_{\VK_{i}}$ ) $\rightarrow$ 0} {$j$ returns $\bot$}
 \Else
 {
$\open$($\VK_{\curi}$, $\VK_{j}$, $lw_{\curi,j}$, $lw_{j,\curi}$) 
 
 }

}
}

\end{algorithm}
\noindent \underline{$\mathsf{Key \ Setup}$, Protocol \ref{algo:keysetup}}: This protocol handles the generation of long-term and temporary identities for all the nodes in the PCN. These keys are used to sign and verify messages in the subsequent protocols of $\ripd$. Initially, all the nodes create their temporary and long-term signing and verification keypairs, denoted by ($\SK$, $\VK$) and ($\sk$, $\vk$) respectively using the $\mathsf{KeyGen}$ function. All the nodes in the PCN send their long-term verification key to their immediate neighbors. Nodes that are immediate neighbors of each other exchange payments in the PCN and hence they need to know each other's real identities.
The temporary verification key of each node is signed using the long-term signing key to produce a signature. This signature ties the long-term identity of a node to its temporary identity.
 This signature is then verified by all the immediate neighbors of a node using the node's long-term verification key. If the signature verifies, the two nodes open a payment channel. The creation of temporary identities is done to hide the real identity of a node in the PCN from its non-neighboring nodes. This helps us achieve our goal of \emph{sender/receiver privacy}.\\
    \begin{algorithm}[htbp]
\caption{$\mathsf{Routing \ Payment}$}
\label{algo:routepay}
\DontPrintSemicolon
\SetAlgoLined
Alice maintains a list $\mathbb{T}$ = $\emptyset$ \\

\For{i =1;i $\leq$ $|\mathbb{P}|$;i++ }
{
Alice performs $\mathbb{T}$.$\add$($\erh_{i}$) \\
}

Alice sends $\mathbb{T}$ to Bob out-of-band and
Bob picks and sends $\erh_{Bob} = \min(\hc_{\erh_{i},Bob})$ $\forall$ $i$ $\in$ $\mathbb{T}$ \\
Alice  calls $\mathsf{ChoosePath}$($\pset$, $\erh_{Bob}$) $\rightarrow$
$\mathbb{P ^ \prime}$\\
 Bob does $X \sample \{0,1\}^\lambda$, $H(X)$ $\rightarrow$ $Y$ and sends $Y$ to Alice \\ 
 \For{each pair of consecutive nodes $i,j$ along the path of $\txid$ from Alice to Bob}
 {

 Alice retrieves the $\txid$ for the transaction to be sent to Bob from the tuple $\mathbb{K}$ = ($\cdot, \txid, \cdot$) \\

 $\curprev$ = $i$, $\curnext$ = $j$ \\
 $\curprev$ sends ($\mathsf{inPath}$, $Y$, $\txid$) to $\curnext$ and \\
 $\curprev$ establishes HTLC with $\curnext$ and
 $\curprev$ = $\curnext$ and $\curnext$ = $\curprev$ +1 \\
 }
\For{every pair of consecutive nodes  along the path of $\txid$ from Bob to Alice }
{

\If{$\curprev$ reveals $X$ to $\curnext$}
{

\If{$\htlcpay$($\vk_{\curprev}$, $\vk_{\curnext}$, $\txid$, $\amount$) $\rightarrow$ $success$ }
{

$\curprev$ = $\curnext$ and $\curnext$ = $\curprev$ -1 \\
}

\Else
{

return $\bot$ \\
}

}
\Else
{

return $\bot$ \\
}

}
\end{algorithm}
 \noindent
\underline{$\mathsf{Routing \ Payment}$, Protocol \ref{algo:routepay}}: This protocol facilitates the routing of payment between Alice and Bob. Initially, Alice retrieves the $\erh$ from each path tuple sent to her in the stack $\mathbb{P}$ by the $\nearrh$ at the end of Protocol~\ref{algo:path1}. The node identifier of the $\erh$ is the last value in each tuple present in $\mathbb{P}$. These $\erh$s are added to the list $\mathbb{T}$ that Alice maintains locally. Alice sends this list to Bob using a secure out-of-band communication channel. Bob picks one $\erh$ that is closest to him based on the minimum hop count between him and the $\erh$.  
Bob notifies Alice regarding his choice of $\erh$ using the same channel. Alice chooses the shortest path that contains the $\rh$ picked by Bob as the $\erh$ using the $\mathsf{ChoosePath}$ function.
This function returns the path, $\mathbb{P}^ \prime$. Bob samples a random pre-image $X$, hashes it to produce a digest $Y$, and sends $Y$ to Alice. For each consecutive node along the path from Alice to Bob, every node sends the tuple ($\mathrm{inPath}$, $Y$, $\txid$) to its immediate neighbor. Upon receiving this tuple, every node along the path establishes an HTLC with its immediate neighbor. Every pair along the path from Bob to Alice reveals the secret used for the HTLC. Upon successful revealing of this secret from every node to its immediate neighbor along the path from Bob to Alice, the payment process is completed. Using HTLCs ensures that no honest party loses any funds because of malicious behavior by other parties in the system, which helps us in achieving our goal of \emph{balance security}. HTLCs also ensure that all the link weights of the nodes along the transaction path go back to the state they were in prior to the commencement of the transaction if the transaction fails for any reason. This achieves our  goal of \emph{atomicity}.

%% file: main.bbl
\begin{thebibliography}{10}

\bibitem{alshaikh}
Moaath Alshaikh and Akram Morie.
\newblock Development of multipath dynamic address routing protocol in manet to improve data transfer in poor infrastructure environment.
\newblock In {\em 2022 International Conference on Computer Science and Software Engineering (CSASE)}, pages 368--373, 2022.

\bibitem{aly2015network}
Abdelrahaman Aly.
\newblock {\em Network flow problems with secure multiparty computation.}
\newblock PhD thesis, Catholic University of Louvain, Louvain-la-Neuve, Belgium, 2015.

\bibitem{aly2013securely}
Abdelrahaman Aly, Edouard Cuvelier, Sophie Mawet, Olivier Pereira, and Mathieu Van~Vyve.
\newblock Securely solving simple combinatorial graph problems.
\newblock In {\em Financial Cryptography and Data Security: 17th International Conference, FC 2013, Okinawa, Japan, April 1-5, 2013, Revised Selected Papers 17}, pages 239--257. Springer, 2013.

\bibitem{abdelrahman}
Abdelrahaman Aly and Mathieu Van~Vyve.
\newblock Securely solving classical network flow problems.
\newblock In Jooyoung Lee and Jongsung Kim, editors, {\em Information Security and Cryptology - ICISC 2014}, pages 205--221, Cham, 2015. Springer International Publishing.

\bibitem{abdelauction}
Abdelrahaman Aly and Mathieu~Van Vyve.
\newblock Practically efficient secure single-commodity multi-market auctions.
\newblock In Jens Grossklags and Bart Preneel, editors, {\em Financial Cryptography and Data Security - 20th International Conference, {FC} 2016, Christ Church, Barbados, February 22-26, 2016, Revised Selected Papers}, volume 9603 of {\em Lecture Notes in Computer Science}, pages 110--129. Springer, 2016.

\bibitem{rippleapi}
Ripple API.
\newblock Ripple api.
\newblock \url{https://data.ripple.com/}.

\bibitem{Hakem}
Hakem Beitollahi and Geert Deconinck.
\newblock Comparing chord, can, and pastry overlay networks for resistance to dos attacks.
\newblock In {\em 2008 Third International Conference on Risks and Security of Internet and Systems}, pages 261--266, 2008.

\bibitem{blockchair}
Blockchair.
\newblock Blockchair.
\newblock \url{https://blockchair.com/ethereum}.

\bibitem{binance}
ETH tx~throughput BTC.
\newblock Btc, eth tx throughput.
\newblock \url{https://academy.binance.com/en/glossary/transactions-per-second-tps}.

\bibitem{DBLP:conf/focs/Canetti01}
Ran Canetti.
\newblock Universally composable security: {A} new paradigm for cryptographic protocols.
\newblock In {\em 42nd Annual Symposium on Foundations of Computer Science, {FOCS} 2001, 14-17 October 2001, Las Vegas, Nevada, {USA}}, pages 136--145. {IEEE} Computer Society, 2001.

\bibitem{canetti2004universally}
Ran Canetti.
\newblock Universally composable signature, certification, and authentication.
\newblock In {\em Proceedings. 17th IEEE Computer Security Foundations Workshop, 2004.}, pages 219--233. IEEE, 2004.

\bibitem{mpcnrp}
Yanjiao Chen, Yuyang Ran, Jingyue Zhou, Jian Zhang, and Xueluan Gong.
\newblock Mpcn-rp: A routing protocol for blockchain-based multi-charge payment channel networks.
\newblock {\em IEEE Transactions on Network and Service Management}, 19(2):1229--1242, 2022.

\bibitem{cormen2022introduction}
Thomas~H Cormen, Charles~E Leiserson, Ronald~L Rivest, and Clifford Stein.
\newblock {\em Introduction to algorithms}.
\newblock MIT press, 2022.

\bibitem{ripplemarketcap}
Ripple current cap.
\newblock Ripple current cap.
\newblock https://www.slickcharts.com/currency.

\bibitem{dasgupta2008algorithms}
Sanjoy Dasgupta, Christos~H Papadimitriou, and Umesh~Virkumar Vazirani.
\newblock {\em Algorithms}.
\newblock McGraw-Hill Higher Education New York, 2008.

\bibitem{Eckey2020SplittingPL}
Lisa Eckey, Sebastian Faust, Kristina Host{\'a}kov{\'a}, and Stefanie Roos.
\newblock Splitting payments locally while routing interdimensionally.
\newblock {\em IACR Cryptol. ePrint Arch.}, 2020:555, 2020.

\bibitem{zelle}
Experian.
\newblock Zelle limit.
\newblock \url{https://bit.ly/49lkpr6}, 2023.

\bibitem{bitkan}
Bitkan Explorer.
\newblock Bitkan explorer.
\newblock \url{https://bit.ly/3LTPiYN}.

\bibitem{lightningfees}
Lightning~Network Fees.
\newblock Lightning network fees.
\newblock \url{https://github.com/lightning/bolts/blob/master/07-routing-gossip.md#htlc-fees}.

\bibitem{xoom}
Forbes.
\newblock xoom.
\newblock \url{https://bit.ly/47iPMkt}, 2023.

\bibitem{goldberg}
A~V Goldberg and R~E Tarjan.
\newblock A new approach to the maximum flow problem.
\newblock In {\em Proceedings of the Eighteenth Annual ACM Symposium on Theory of Computing}, STOC '86, page 136–146, New York, NY, USA, 1986. Association for Computing Machinery.

\bibitem{vein}
Qianyun Gong, Chengjin Zhou, Le~Qi, Jianbin Li, Jianzhong Zhang, and Jingdong Xu.
\newblock Vein: High scalability routing algorithm for blockchain-based payment channel networks.
\newblock In {\em 2021 IEEE 20th International Conference on Trust, Security and Privacy in Computing and Communications (TrustCom)}, pages 43--50, 2021.

\bibitem{auto}
Hsiang-Jen Hong, Sang-Yoon Chang, and Xiaobo Zhou.
\newblock Auto-tune: Efficient autonomous routing for payment channel networks.
\newblock In {\em 2022 IEEE 47th Conference on Local Computer Networks (LCN)}, pages 347--350, 2022.

\bibitem{htlc}
LND HTLC.
\newblock Lnd htlc.
\newblock \url{https://docs.lightning.engineering/the-lightning-network/multihop-payments/hash-time-lock-contract-htlc}.

\bibitem{machine}
Heba Kadry and Yasser Gadallah.
\newblock A machine learning-based routing technique for off-chain transactions in payment channel networks.
\newblock In {\em 2021 IEEE International Conference on Smart Internet of Things (SmartIoT)}, pages 66--73, 2021.

\bibitem{Keypair}
LND keypair.
\newblock Lnd keypair.
\newblock \url{https://github.com/lightning/bolts/blob/master/08-transport.md}.

\bibitem{networkx}
Networkx library.
\newblock Networkx library.
\newblock \url{Phttps://networkx.org/}.

\bibitem{FSTR}
Siyi Lin, Jingjing Zhang, and Weigang Wu.
\newblock Fstr: Funds skewness aware transaction routing for payment channel networks.
\newblock In {\em 2020 50th Annual IEEE/IFIP International Conference on Dependable Systems and Networks (DSN)}, pages 464--475, 2020.

\bibitem{malavolta2017concurrency}
Giulio Malavolta, Pedro Moreno-Sanchez, Aniket Kate, Matteo Maffei, and Srivatsan Ravi.
\newblock Concurrency and privacy with payment-channel networks.
\newblock In {\em Proceedings of the 2017 ACM SIGSAC Conference on Computer and Communications Security}, pages 455--471, 2017.

\bibitem{Malavolta2016SilentWhispersES}
Giulio Malavolta, Pedro~A. Moreno-Sanchez, Aniket Kate, and Matteo Maffei.
\newblock Silentwhispers: Enforcing security and privacy in decentralized credit networks.
\newblock {\em IACR Cryptol. ePrint Arch.}, 2016:1054, 2016.

\bibitem{Cryptostats}
BTC market cap.
\newblock Btc market cap.
\newblock \url{https://coinmarketcap.com/currencies/bitcoin/}.

\bibitem{ripplemarket}
Ripple market value.
\newblock Ripple market capitalization.
\newblock \url{https://bit.ly/3AGVnT0}.

\bibitem{maymounkov2002kademlia}
Petar Maymounkov and David Mazieres.
\newblock Kademlia: A peer-to-peer information system based on the xor metric.
\newblock In {\em International Workshop on Peer-to-Peer Systems}, pages 53--65. Springer, 2002.

\bibitem{message}
LND message passing.
\newblock Lnd message passing.
\newblock \url{https://github.com/lightning/bolts/blob/master/07-routing-gossip.md}.

\bibitem{rpca}
Ripple message passing.
\newblock Ripple message passing.
\newblock \url{https://ripple.com/files/ripple_consensus_whitepaper.pdf}.

\bibitem{sprites}
Andrew Miller, Iddo Bentov, Ranjit Kumaresan, and Patrick McCorry.
\newblock Sprites: Payment channels that go faster than lightning.
\newblock 02 2017.

\bibitem{minlightning}
LN~minimum payment.
\newblock Ln minimum payment.
\newblock \url{https://dci.mit.edu/lightning-network}.

\bibitem{monfared2023darvan}
Saleh~Khalaj Monfared and Saeed Shokrollahi.
\newblock Darvan: A fully decentralized anonymous and reliable routing for vanets.
\newblock {\em Computer Networks}, 223:109561, 2023.

\bibitem{nakamoto2008bitcoin}
Satoshi Nakamoto.
\newblock Bitcoin: A peer-to-peer electronic cash system.
\newblock {\em Decentralized Business Review}, page 21260, 2008.

\bibitem{xoomlimit}
Nerdwallet.
\newblock Xoom limit minimum.
\newblock \url{https://bit.ly/3QnPOAu}, 2023.

\bibitem{lightning}
Lightning Network.
\newblock Lightning network.
\newblock \url{https://lightning.network/}.

\bibitem{stellar}
Stellar Network.
\newblock Stellar network.
\newblock \url{https://www.stellar.org/?locale=en}.

\bibitem{blanc}
Gaurav Panwar, Satyajayant Misra, and Roopa Vishwanathan.
\newblock Blanc: Blockchain-based anonymous and decentralized credit networks.
\newblock In {\em Proceedings of the Ninth ACM Conference on Data and Application Security and Privacy}, pages 339--350, 2019.

\bibitem{flare}
Flare PCN.
\newblock Flare.
\newblock \url{https://flare.xyz/the-flare-network/}.

\bibitem{lightpir}
Krzysztof Pietrzak, Iosif Salem, Stefan Schmid, and Michelle Yeo.
\newblock Lightpir: Privacy-preserving route discovery for payment channel networks.
\newblock In {\em 2021 IFIP Networking Conference (IFIP Networking)}, pages 1--9, 2021.

\bibitem{pool}
Lightning Pool.
\newblock Lightning pool.
\newblock \url{https://lightning.engineering/lightning-pool-whitepaper.pdf}.

\bibitem{ripple}
Ripple.
\newblock Ripple.
\newblock \url{https://ripple.com/}.

\bibitem{roos2016voute}
Stefanie Roos, Martin Beck, and Thorsten Strufe.
\newblock Voute-virtual overlays using tree embeddings.
\newblock {\em arXiv preprint arXiv:1601.06119}, 2016.

\bibitem{roos2017settling}
Stefanie Roos, Pedro Moreno{-}Sanchez, Aniket Kate, and Ian Goldberg.
\newblock Settling payments fast and private: Efficient decentralized routing for path-based transactions.
\newblock In {\em 25th Annual Network and Distributed System Security Symposium, {NDSS} 2018, San Diego, California, USA, February 18-21, 2018}. The Internet Society, 2018.

\bibitem{rowstron2001pastry}
Antony Rowstron and Peter Druschel.
\newblock Pastry: Scalable, decentralized object location, and routing for large-scale peer-to-peer systems.
\newblock In {\em IFIP/ACM International Conference on Distributed Systems Platforms and Open Distributed Processing}, pages 329--350. Springer, 2001.

\bibitem{btcpay}
BTCPAY Server.
\newblock Btcpay server.
\newblock \url{https://bit.ly/3q3oAlU}.

\bibitem{sivaraman2020high}
Vibhaalakshmi Sivaraman, Shaileshh~Bojja Venkatakrishnan, Kathy Ruan, Parimarjan Negi, Lei Yang, Radhika Mittal, Mohammad Alizadeh, and Giulia Fanti.
\newblock High throughput cryptocurrency routing in payment channel networks, 2020.

\bibitem{srivatsa}
Mudhakar Srivatsa and Ling Liu.
\newblock Mitigating denial-of-service attacks on the chord overlay network: A location hiding approach.
\newblock {\em IEEE Transactions on Parallel and Distributed Systems}, 20(4):512--527, 2009.

\bibitem{chord}
Ion Stoica, Robert Morris, David Karger, M.~Frans Kaashoek, and Hari Balakrishnan.
\newblock Chord: A scalable peer-to-peer lookup service for internet applications.
\newblock In {\em Proceedings of the 2001 Conference on Applications, Technologies, Architectures, and Protocols for Computer Communications}, SIGCOMM '01, page 149–160, New York, NY, USA, 2001. Association for Computing Machinery.

\bibitem{subramanian2020balance}
Lalitha~Muthu Subramanian, Roopa Vishwanathan, and Kartick Kolachala.
\newblock Balance transfers and bailouts in credit networks using blockchains.
\newblock In {\em 2020 IEEE International Conference on Blockchain and Cryptocurrency (ICBC)}, pages 1--3. IEEE, 2020.

\bibitem{tran}
Hieu Tran, Miao Miao, Farokh Bastani, and I-Ling Yen.
\newblock Multi-keyword based information routing in peer-to-peer networks.
\newblock In {\em 2023 International Conference on Information Networking (ICOIN)}, pages 791--796, 2023.

\bibitem{visa}
VISA transactions.
\newblock Visa transactions.
\newblock \url{https://www.visa.co.uk/dam/VCOM/download/corporate/media/visanet-technology/aboutvisafactsheet.pdf}.

\bibitem{zied}
Zied Trifa.
\newblock Preventing sybil attacks in chord and kademlia protocols.
\newblock {\em International Journal of Internet Protocol Technology}, 12(3):157--166, 2019.

\bibitem{trustlines}
Ripple trustline API.
\newblock Ripple trustline api.
\newblock \url{https://xrpl.org/account\_lines.html}.

\bibitem{btctx}
BTC tx~throughput.
\newblock Btc tx throughput.
\newblock \url{https://www.blockchain.com/explorer/charts/transactions-per-second}.

\bibitem{flash}
Peng Wang, Hong Xu, Xin Jin, and Tao Wang.
\newblock Flash: Efficient dynamic routing for offchain networks.
\newblock CoNEXT '19, page 370–381, New York, NY, USA, 2019. Association for Computing Machinery.

\bibitem{coinexpress}
Ruozhou Yu, Guoliang Xue, Vishnu Kilari, Dejun Yang, and Jian Tang.
\newblock Coinexpress: A fast payment routing mechanism in blockchain-based payment channel networks.
\newblock pages 1--9, 07 2018.

\bibitem{zahid2022fault}
Saleem Zahid, Kifayat Ullah, Abdul Waheed, Sadia Basar, Mahdi Zareei, and Rajesh~Roshan Biswal.
\newblock Fault tolerant dht-based routing in manet.
\newblock {\em Sensors}, 22(11):4280, 2022.

\bibitem{webflow}
Xiaoxue Zhang, Shouqian Shi, and Chen Qian.
\newblock Webflow: Scalable and decentralized routing for payment channel networks with high resource utilization.
\newblock {\em CoRR}, abs/2109.11665, 2021.

\bibitem{robustpay}
Yuhui Zhang and Dejun Yang.
\newblock Robustpay: Robust payment routing protocol in blockchain-based payment channel networks.
\newblock In {\em 2019 IEEE 27th International Conference on Network Protocols (ICNP)}, pages 1--4, 2019.

\bibitem{robustpay++}
Yuhui Zhang and Dejun Yang.
\newblock Robustpay+: Robust payment routing with approximation guarantee in blockchain-based payment channel networks.
\newblock {\em IEEE/ACM Transactions on Networking}, 29(4):1676--1686, 2021.

\bibitem{zhao2004tapestry}
Ben~Y Zhao, Ling Huang, Jeremy Stribling, Sean~C Rhea, Anthony~D Joseph, and John~D Kubiatowicz.
\newblock Tapestry: A resilient global-scale overlay for service deployment.
\newblock {\em IEEE Journal on selected areas in communications}, 22(1):41--53, 2004.

\end{thebibliography}
